\documentclass[11pt]{amsart}
\usepackage{amsfonts,amsmath,amssymb,amsthm}
\usepackage{latexsym}
\usepackage{eucal}
\usepackage[dvips]{graphics}
\usepackage[english]{babel}
\usepackage{epsfig}

\newcommand{\nwc}{\newcommand}
\nwc\eps{\varepsilon}

\nwc{\be}{\begin{equation}}
\nwc{\ee}{\end{equation}}
\nwc{\bee}{\begin{eqnarray}}
\nwc{\eee}{\end{eqnarray}}
\nwc{\ba}{\begin{array}}
\nwc{\ea}{\end{array}}

\nwc{\D}{\partial}
\nwc{\ip}[1]{{\langle #1 \rangle}}
\nwc{\ipbig}[1]{{\left\langle #1 \right\rangle}}
\nwc{\Span}[1]{\mathop{\rm span}\nolimits\{#1\}}
\nwc{\sfrac}[2]{{\textstyle\frac{ #1}{ #2}}}

\newcommand\ep{{\epsilon}}

\newcommand\nd{\noindent}

\newcommand\dx{\partial_x}
\newcommand\V{{\mathcal{V}}}
\newcommand\W{{\mathcal{W}}}
\newcommand\A{{\mathcal{A}}}

\newcommand\IH{{\tilde{H}}}
\newcommand\IL{{\mathcal{L}}}

\newcommand\R{\mathbb{R}}
\newcommand\C{\mathbb{C}}

\newcommand\Z{\mathbb{Z}}

\newcommand\E{{\mathcal{E}}}
\newcommand\Ra{{\mathcal{R}}}
\newcommand\Pa{{\mathcal{P}}}

\newcommand\la{\lambda}
\newcommand\ga{\gamma}
\newcommand\Ga{\Gamma}

\newcommand\bt{\beta}

\nwc{\bond}{\sigma}

\newcommand{\md}{{\mathcal M}_\delta}

\newcommand{\usp}{{\US_1(n)+\US_2(n)}}
\newcommand{\usm}{{\US_1(n)-\US_2(n)}}

\nwc{\CIH}[1]{\C \tilde H^{#1}}

\nwc{\jj}{{m}}
\nwc{\US}{U^{\#}}
\nwc{\GS}{G^{\#}}

\nwc{\phd}{\phi_\delta}
\nwc{\bo}{\sigma}
\nwc{\fdel}{f_\delta}
\nwc{\Fd}{F_\delta}
\nwc{\Fdd}{\tilde F_\delta}
\nwc{\Forall}{\mbox{ for all\ }}
\nwc{\Sc}{S_0}
\nwc{\Sh}{S_1}
\nwc{\Mt}{\tilde{M}(\beta)}
\nwc{\umap}{\tilde{u}}

\nwc{\CM}{\widetilde{\cal M}_\delta}
\nwc{\Lip}{{\rm Lip}}
\nwc{\pibd}{c_\pi}
\nwc{\fix}{T}
\nwc{\tfix}{\tilde{T}}

\newcommand{\Ronetwo}{{\mathcal R}} 
\newcommand{\Ronetwoinv}{{\mathcal P}} 
\newcommand{\wtoU}{\phi}
\newcommand{\HHone}{{\IH^1\times\IH^1}}
\newcommand{\HHtwo}{{\IH^2\times\IH^2}}
\newcommand{\vonetwo}{\tilde{V}}
\newcommand{\ponetwo}{\tilde{\phi}}
\newcommand{\Aonetwo}{{\mathcal A}}

\nwc{\SH}{S_{H}}
\newtheorem{Def}{Definition}[section]
\newtheorem{theorem}{Theorem}[section]
\newtheorem{corollary}{Corollary}[section]

\newtheorem{lemma}{Lemma}[section]

\thanks{Research  supported by Universidad del Valle, Cali- Colombia}
\email{jose.quintero@correounivalle.edu.co}

\begin{document}
\title[A center manifold application: travelling  waves existence]{\bf A center manifold application: existence of periodic travelling waves for the 2D $abcd$-Boussinesq system}
\author[J.R. Quintero]{Jos\'{e} R. Quintero, Alex Montes}
\maketitle

\begin{abstract} In this paper we study  the existence of periodic travelling waves for  the  2D $abcd$ Boussinesq type system related with the  three-dimensional water-wave dynamics in the weakly nonlinear long-wave regime. We show that small solutions that are periodic in the direction of translation (or orthogonal to it) form an infinite-dimensional family, by characterizing these solutions   through spatial dynamics when we reduce to a center  manifold of infinite dimension and codimension the linearly ill-posed mixed-type initial-value problem. As happens for the Benney-Luke model and the KP II model for wave speed large enough and large surface tension, we show that a unique global solution exists for arbitrary small initial data for the two-component bottom velocity, specified along a single line in the direction of translation (or orthogonal to it). As a consequence of this fact, we show that the spatial evolution of bottom velocity is governed by a dispersive, nonlocal, nonlinear wave equation.
\end{abstract}


\nd {\bf keywords: periodic travelling waves, center manifold approach, stability}

\section{\bf Introduction}

In this paper we show the existence of travelling waves  for a nonlinear  water waves system which are periodic in their direction of propagation and have a pulse structure in the unbounded transverse spatial direction. We develop an approach to describing  steady water waves
through spatial dynamics and center manifold reduction  to obtain such a result for a 2D  water-wave system, in which the expression  ``spatial dynamics" means to perform a method in such  that a  system of partial differential equations governing is formulated as an
evolutionary equation (in general ill-posed)
\begin{equation} \label{evol}
u_{\zeta} = A(u) + G(u),
\end{equation}
where an unbounded spatial coordinate plays the role of the timelike variable $\zeta$ (see Kirchg\"{a}issner \cite{Ki82}). For the Boussinesq  system considered in this work,  we are able to describe all small waves
that translate steadily with supercritical speed and  that are periodic in the direction of translation (or orthogonal to it). Moreover, these solutions  form an infinite-dimensional
family which can be parametrized by the bottom-velocity profile along any single line in the direction of translation (or orthogonal to it), meaning that for a fixed  supercritical wave speed and an arbitrary bottom-velocity profile small in a suitable Sobolev space, there
corresponds a unique globally defined periodic travelling. As we know, in the linear wave equation for the surface elevation $\eta$ (nondimensional)
$$
\eta_{tt} = \eta_{xx}+\eta_{yy},
$$
is possible to have  many traveling-wave solutions $\eta=f(x-ct,y)$ translating with supercritical speed $|c|>1$, by solving the wave equation $f_{yy} =(c^2-1)f_{\xi\xi}$ with given arbitrary initial data for the wave slope $(\eta_x,\eta_y)$ along the single line $y=0$. Moreover, for the exact linearized water wave equations for an inviscid irrotational fluid without surface tension, and  for the Kadomtsev-Petviashvili equation in the KP-II case, we know that for arbitrary small data for wave slope along a line determine a unique travelling wave with given supercritical speed (see \cite{HP}). On the other hand, J. Quintero and R. Pego in (\cite{PQ2}) showed that  a complete periodic travelling wave solution for the Benney-Luke model is determined uniquely by arbitrarily specifying along $y=0$ the horizontal velocity $(\Phi_x,\Phi_y)$ at the fluid bottom, provided these data are sufficiently small in an appropriate norm (the $H^1$ Sobolev norm).

In this paper we consider the existence of non trivial periodic travelling
wave for the 2D $abcd$-Boussinesq  system related with the water wave problem
\begin{equation}
\left\{
\begin{array}{r}
(I- b\mu\Delta)\Phi _{t}+ (I- a\mu\Delta)\eta
+\frac{\ep}{p+1}\left( \Phi _{x}^{p+1}+\Phi _{y}^{p+1}\right) =0, \\
\\
(I- d\mu\Delta)\eta _{t} +\Delta \Phi -c\mu\Delta ^{2}\Phi+ \ep \nabla \cdot \left( \eta \left( \Phi _{x}^{p},\Phi
_{y}^{p}\right) \right) =0,
\end{array}
\right.   \label{eq:bts1}
\end{equation}
where $b+d-a-c=\frac13-\sigma$, with $a, b, c, d >0$,  $\sqrt \mu ={h_{0}}/{L}$ is the ratio of undisturbed fluid depth to typical wave length (long-wave parameter or dispersion coefficient), and $\epsilon $ is the ratio of typical wave amplitude to fluid depth (amplitude parameter or nonlinearity coefficient), $\sigma$ is associated with the surface tension ($\sigma ^{-1}$ is the Bond number), $\Phi $ is the rescale nondimensional velocity potential on the bottom $z=0$, and $\eta $  is the rescaled free surface elevation. We consider waves which are periodic in a moving frame of reference, so that they are periodic in the variable $x - ct$, where $t$ denotes the temporal variable. For this physical problem, we have a bounded spacelike coordinates (the vertical direction), which is bounded because the fluid has finite depth, and the coordinate $x - ct$ which is considered bounded because we are looking for periodic wave in this variable. So, since there is not any  restriction upon the behavior of the waves in the spatial direction $y$ transverse
to their direction of propagation, we are allowed to use $y$ as timelike variable. We apply spatial
dynamics to the problem by formulating this  as an evolutionary system of the form (\ref{evol}) with $\zeta=y$ in an infinite-dimensional phase space consisting of periodic functions of $x - ct$ (see  Groves and Schneider \cite{GS2001}, Sandstede and Scheel \cite{SS1999}, \cite{SS2004}, and Haragus-Courcelle and Schneider \cite{HS1999} for applications to respectively nonlinear wave equations, reaction-diffusion equations, and Taylor-Couette problems, Quintero and Pego \cite{PQ2} for periodic nonlinear travelling for the Benney-Luke model).

In the case of wave speed $|\omega|>1$ and  $\sigma>\frac12$ (large surface tension), the $abcd$-Boussinesq system (\ref{eq:bts1}) under consideration   has a very close relationship with the Benney-Luke model derived by J. Quintero and R. Pego in \cite{QP1} when  $\tilde a- \tilde b= \sigma-\frac13>0$ and also with the KP-II model ($\sigma>\frac13$), in the sense that travelling waves for the Boussinesq system (\ref{eq:bts1}) can generate travelling waves for the Benney-Luke model and the KP-II model, up to some order. Moreover, for large wave speed and large surface tension, the $abcd$-Boussinesq system (\ref{eq:bts1}), the Benney-Luke model and the KP II  model share ill-posed spatial evolution equations with infinite-dimensional center manifolds, where  the dimension of the center manifold is determined by the number of purely imaginary eigenvalues (e.g., see Vanderbauwhede \cite{V1989}). Existence of invariant center manifold of infinite dimension and codimension under special hypotheses has been established  for infinite-dimensional evolutionary systems, showing in particular that  the original problem is locally equivalent to a system of ordinary differential equations whose solutions can be expressed in terms of the solution on the center space (tangent to the center manifold), see for example works  by Scarpellini \cite{Sc90}, Mielke \cite{M1991} and Vanderbauwhede and  Iooss \cite{VI89}, Quintero and Pego \cite{PQ2}, among others. In the $abcd$-Boussinesq, existence of an infinite dimensional invariante center manilfod for which the spectrum of hyperbolic space  is unbounded on both sides of the imaginary axis is a direct consequence of the work by J. Quintero and R. Pego \cite{PQ2} (see also \cite{Ki82,Mi88a,Mi92}).

In the work we describe all small travelling waves that translate steadily, which  are periodic in the direction of translation (or orthogonal to it), in the case of  supercritical speed $|\omega|>1$ and some restriction on the coefficients $a, b, c, d$. In this regime, after rescaling $\epsilon$ and $\mu $, the traveling-wave system for (\ref{eq:bts1}) takes the form
\begin{equation}
\left(
\begin{array}{c}
 -\omega(I-d\Delta)u_{x} + (I-c\Delta)\Delta v+\nabla \cdot \left( u\left(
v_{x}^{p},v_{y}^{p}\right) \right) \\
 -\omega(I-b\Delta)v_{x}+(I-a\Delta)u+\frac{1}{p+1}\left(
v_{x}^{p+1}+v_{y}^{p+1}\right)
\end{array}
\right) =\left(
\begin{array}{c}
0 \\
0
\end{array}
\right).   \label{eq:travel}
\end{equation}
Existence of periodic travelling waves follows by considering the system (\ref{eq:travel}) as an evolution equation likewise (\ref{evol}), where $y$ acts as the ``time'' variable.  If we seek for $x$-periodic travelling wave solutions, the initial-value problem for the system (\ref{eq:travel}) considered as an evolution equation in the variable $y$ has mixed type due to the fact that the  Cauchy problem turns out to be linearly \emph{ill-posed} for wave speed $|\omega|$ large enough and under some conditions on $a, b, c$ and $d$. We will see that at  the linear level there are infinite many central modes (pure imaginary eigenvalues) and infinitely many hyperbolic modes, and also that there is a spectral gap since the hyperbolic nodes are located outside of a strip containing the imaginary axis. As a consequence of this fact, the existence result of $x$-periodic travelling wave solutions involves using an invariant center manifold of infinite dimension and infinite codimension. This center manifold contains all globally defined small-amplitude solutions of the travelling wave equation for the Boussinesq system that are $x$-periodic in the direction of propagation.

We point out that the dynamics for the $abcd$-Boussinesq system in the case $a, c>0$ and $b=d=0$ is quite different from the case   $a, b, c, d >0$ in the sense that in the first case, at the linear level,  there are only finite many central modes (pure imaginary eigenvalues) and infinitely many
hyperbolic modes (see \cite{JQ-2015RACC}).

The main result of this paper involves using the fact that there exists an invariant
center manifold of infinite dimension and codimension that contains all globally defined small-amplitude solutions of \eqref{eq:travel} that are periodic in the direction of propagation.  These solutions are determined by suitable initial data along the line $y=0$. In particular,
we show that a complete traveling wave solution of \eqref{eq:travel} is determined uniquely by arbitrarily specifying along $y=0$ the horizontal velocity $(\Phi_x,\Phi_y)$ at the fluid bottom, provided these data are sufficiently small in an appropriate norm (the $H^1$ Sobolev norm).
The bottom-velocity profile evolves in $y$ according to a dispersive, nonlinear, nonlocal wave equation obtained by restriction to the two first components  to the invariant center manifold of the form
\[
\frac{d}{dy}\begin{pmatrix}u_1\cr u_2\end{pmatrix} =
\begin{pmatrix}0&\dx\cr S\dx&0\end{pmatrix}\begin{pmatrix}u_1\cr u_2\end{pmatrix}+
\begin{pmatrix}0\cr g(u_1,u_2)\end{pmatrix}.
\]
in which the map
$g$ is Lipschitz with $g(0)=0$, $Dg(0,0)=0$, and $S$ is a nonlocal
linear operator of zero order.

\section{\bf Existence of a local center manifold}
The goal in this section is to verify the hypotheses of an abstract center manifold result obtained by J. Quintero and R. Pego in (\cite{PQ2}) (Theorem \ref{T:1} in the Appendix)   to get  the existence of a local center manifold associated with travelling waves for  $abcd$-Boussinesq system, where the  is viewed as a first order system \eqref{eq:stw} with the variable $y$ being the timelike variable. We will see  for the system  \eqref{eq:stw} associated with the  $abcd$-Boussinesq system, that there exists a locally invariant center manifold of classical solutions, where the center subspace (that associated with the
purely imaginary spectrum of $A$)  has infinite dimension and
infinite  codimension, as happens for  the Benney-Luke model (see \cite{PQ2}).

Hereafter, for a given integer $r\geq 0$, let $\tilde H^r$ denote the Sobolev space of $2\pi$-periodic functions on $\R$ whose weak derivatives up to order $k$ are square-integrable with norm given by
\[
\|u\|_{\IH^r}^2 = \sum_{j=0}^r \int_0^{2\pi} |\dx^j u|^2\,dx
\]
We will study the existence of $x$-periodic solutions for (\ref{eq:stw}) in the Hilbert spaces $H$ and $X$ defined by
\begin{eqnarray}
H&=& \IH^{1} \times \IH^{1} \times \IH^{0} \times \IH^{-1}\times \IH^{1} \times \IH^{0}, \\
X&=& \IH^{2} \times \IH^{2} \times \IH^{1} \times \IH^{0}\times \IH^{1} \times \IH^{0} .
\end{eqnarray}
Note that $X$ is densely embedded in $H$.
\subsection{\bf Periodic travelling wave equation for the $abcd$ Boussinesq system}

Recall that $x$-periodic travelling-wave profile $(u, v)$ should satisfy the system (\ref{eq:travel}). In order to look for the existence of $x$-periodic travelling waves of period $2\pi$, we set the new variables
\[
u_1 = \partial_xv,  u_2 =  \partial_yv,  u_3 =  \partial_{yy}v,  u_4 = \partial_{yyy}v,  u_5 = u,  u_6 = \partial_yu,
\]
then we have for $\gamma c=1$ and $a c \alpha=1$  that
\begin{align}
\partial_yu_1 &= \partial_xu_2 \label{du1}\\
\partial_yu_2 &= u_3 \label{du2}\\
\partial_yu_2 &= u_4 \label{du3}\\
\partial_y u_4&= -\left(\alpha d\omega^2-\gamma\right)\partial_x u_1+ (\alpha bd \omega^2-1)\partial_{xxx} u_1
+ \gamma u_3 +(\alpha bd \omega^2-2)\partial_{xx} u_3 \nonumber\\
& -\omega (\gamma-\alpha d) \partial_xu_5  +\gamma(u_5u_1^p)_x +\gamma(u_6 u_2^p+ p u_5u_2^{p-1} u_3 ) +\frac{\alpha d \omega }{p+1}\partial_x\left(u_1^{p+1}+u_2^{p+1}\right)\label{du4}  \\
\partial_y u_5 &= u_6 \label{du5}\\
\partial_y u_6&= -\omega c \alpha  u_1+ {b\omega c \alpha} \partial_{xx} u_1+{b\omega c \alpha}\partial_x u_3
+\alpha cu_5- \partial_{xx}u_5+\frac{\alpha c}{p+1}\left(u_1^{p+1}+u_2^{p+1}\right) \label{du6}.
\end{align}
In terms of the new variable $U^t =(u_1,  u_2,  u_3,  u_4, u_5, u_6)$, we see that this system can be rewritten as an evolution in which $y$ is considered as the time variable
\begin{equation}\partial_yU = AU + G(U), \label{eq:stw}
\end{equation}
where we have that
\begin{align*}
\A &= \begin{pmatrix}
0 & \partial_x & 0 & 0 & 0 &0\\
0& 0 & I&  0 & 0&  0 \\
0& 0& 0 &I & 0 & 0\\
-\left(\alpha d\omega^2-\gamma \right)\partial_x + (\alpha bd \omega^2-1)\partial_{xxx} & 0& \gamma I +(\alpha bd \omega^2-2)\partial_{xx} &0& -\omega ( \gamma-\alpha d) \partial_x& 0\\
0& 0& 0 & 0 & 0 &I \\
-\omega c \alpha  I + {b\omega c \alpha} \partial_{xx} &  0&
{b\omega c \alpha}\partial_x   & 0 &c\alpha I- \partial_{xx} &0
\end{pmatrix}
\intertext{and also}
G(U) &= \begin{pmatrix}
0 \\
0\\
0\\
\gamma(u_5u_1^p)_x +\gamma(u_6 u_2^p+ p u_5u_2^{p-1} u_3 ) +
\frac{\alpha d \omega }{p+1}\partial_x\left(u_1^{p+1}+u_2^{p+1}\right) \\
0\\
\frac{\alpha c }{p + 1}\left(u_1^{p+1} + u_2^{p+1}\right)
\end{pmatrix}
\end{align*}
 If we assume that $U(x; y) = \underset{n\in \Z}\sum \widehat U
(n,  y)e^{inx}$, then we see that
\[
\partial_y \widehat U(n)= \widehat{\A}(n) \widehat{U}(n, y) + \widehat{G}_n(U),
\]
where the matrix $\widehat{\A}(n)$ has the form
\[
\widehat{\A}(n) = \begin{pmatrix}
0 & in  & 0 & 0 & 0 &0\\
0& 0 & 1&  0 & 0&  0 \\
0& 0& 0 &1& 0 & 0\\
-in \left(\alpha d\omega^2-\gamma \right) -in^3(\alpha bd \omega^2-1) & 0& \gamma  - n^2(\alpha bd \omega^2-2) &0& -in \omega (\gamma-\alpha d) & 0\\
0& 0& 0 & 0 & 0 &1 \\
-\omega c \alpha   -n^2{b\omega c \alpha} &  0& in{b\omega c \alpha}
& 0 &c\alpha +n^2 &0
\end{pmatrix}
\]
It is straightforward to see that the characteristic polynomial $\mathcal P(n, \beta)$ of $\widehat{\A}(n)$ is given by
\begin{equation}
\mathcal P(n, \beta)= \beta^6+a_2(n)\beta^4+a_1(n)\beta^2
+a_0(n), \label{be}
\end{equation}
where $a_0(n)$, $a_1(n)$ and $a_2(n)$ are defined as
\begin{align}
a_2(n)&=(\alpha b d \omega ^2 -3)n^2 -(c\alpha + \gamma) \label{a2}\\
a_1(n) &= (3 -2\alpha bd \omega ^2 )n^4 + (2(c\alpha +\gamma) -
\alpha \omega ^2(b+d))n^2
+ \alpha \label{a1}\\
a_0(n) &=n^2 \left[(\alpha b d \omega ^2 -1)n^4 + (\alpha
(d\omega^2-c) + \gamma( b c \alpha \omega^2  -1))n^2 +
\left(\alpha(\omega^2-1)\right)\right] \label{a0}
\end{align}
We note that the eigenvalues for $\widehat{\A}(n)$ are roots of the cubic polynomial $p_n$ in the variable $\lambda = \beta^2$ given by
\begin{equation}
p_n(\la)= \lambda^3+a_2(n)\la^2+a_1(n)\la
+a_0(n), \label{la}
\end{equation}

\subsection{Structural conditions}

First consider the basic structural conditions. It is straightforward
to check that $A\in\IL(X,H)$, the space of bounded linear operators from $X$ to $H$.
It is easy to check for $ U, V \in H$ that
\be
\|f(U+V) -f(U)\|_{X} \le (\|U\|^p_H+\|V\|^p_H)\|V\|_H.
\label{fest}
\ee
In other words, $f\colon H\to X$ is smooth, and clearly $f(0)=0=Df(0)$, meaning that
the nonlinear map $f$ exhibits a gain of regularity. We will exploit this feature
to establish the existence of classical solutions.

\subsection{Spectral decomposition}

We will construct the desired spectral decompositions
of $H$ and $X$ by using a complete set of eigenfunctions built via Fourier transform. First, we see that any element $U\in H$ or $X$ can be represented by a Fourier series
\be
U= \sum_{n \in \Z}
\hat{U}(n)e^{inx}.
\label{3-uhat}
\ee
In terms of the vector Fourier coefficients, the norms in $H$ and $X$ may be given by
\be\label{FTnorm}
\|U\|^{2}_{H} = \sum_{n \in \Z}|\SH(n) \hat{U}(n)|^2,
\qquad
\|U\|^{2}_{X} = \sum_{n \in \Z}(1+n^2)|\SH(n) \hat{U}(n)|^2,
\ee
where
\[
\SH(n)= diag \left\{(1+n^2)^{1/2}, (1+n^2)^{1/2}, 1,
(1+n^2)^{-1/2}, (1+n^2)^{1/2}, 1\right\}.
\]

\subsection*{\bf Right and left eigenvectors of $\widehat \A(n)$}
For $n=0$, we have that
\[
p_0(\la)= \lambda^3-  (c\alpha+\gamma)\la^2+c\alpha\gamma\la=\la(\la - c\alpha)(\la-\gamma).
\]
So, we conclude that $\la=0$, $\la=c\alpha$ and  $\la=\gamma$. It
can be seen that $\beta_1(0) = \beta_2(0) = 0$ is a double zero of
$\mathcal P(0, \beta)$, and $\beta_3(0)= \sqrt{\ga} = -\beta_4(0)$
and $\beta_5(0)= \sqrt{c\alpha} = -\beta_6(0)$ are simple roots of
$\mathcal P(0, \beta)$.

We see directly that the eigenvectors are
\[
v_1(0) = (1, 0, 0, 0, \omega, 0)^t,\ \ \ v_2(0) = (0, 1, 0, 0, 0,
0)^t.
\]
and left eigenvectors
\[
z_1(0) = (1, 0, 0, 0, 0, 0)^t, \ \ \  z_2(0) = (0, 1, 0, -c,
0, 0)^t.
\]
The eigenvalue $\beta_3(0)= \sqrt{\ga} = -\beta_4(0)$ are single
eigenvalues with right eigenvectors
\[
v_3(0) = \left(0, 1, \sqrt{\frac1{c}}, \frac1{c},  0, 0\right)^t,\ \ \ v_4(0) = \left(0, 1,
-\sqrt{\frac1{c}}, \frac1{c},  0, 0\right)^t.
\]
and left eigenvectors
\[
z_3(0) =\frac{c}{2} \left(0, 0,  \sqrt{\frac1{c}}, 1,0, 0\right)^t,\ \ \ z_4(0) =\frac{c}{2}\left(0, 0, -\sqrt{\frac1{c}}, 1,0, 0\right)^t.
\]
On the other hand, the eigenvalue $\beta_5(0)= \sqrt{c\alpha} =
-\beta_6(0)$ are single eigenvalues with right eigenvectors
\[
v_5(0) = \left(0, 0, 0, 0, 1, \sqrt{\frac1{a}}\right)^t,\ \ \ v_6(0) = \left(0, 0, 0,
0, 1, -\sqrt{\frac1{a}}\right)^t.
\]
and left eigenvectors

\[
z_5(0) = \frac{\omega}{2}\left(1, 0, 0, 0, \frac{1}{\omega},
\frac{\sqrt{a}}{\omega}\right)^t,\ \ \ z_6(0) = \frac{\omega}{2}\left(1, 0, 0, 0, \frac{1}{\omega}, \frac{-\sqrt{a}}{\omega}\right)^t.
\]

Assume now  that $n \ne  0$.  We will see that the polynomial $p_n$
has a  negative real root $\lambda_1(n)$ for any $n \ne 0$ for
$\omega^2> \max \left\{1, \frac{c}{d}, \frac{1}{ \alpha bd},
\frac{1}{ \alpha bc} \right\} = \max \left\{1, \frac{c}{d}, \frac{ac}{bd},
\frac{a}{ b} \right\}:= \omega_0^2(a, b, c, d)$ with
$w_0>0$ ($b, d>0)$. In fact, in this regime, we have that $a_0(n) >0$ for $n
\ne 0$ and so, $p_n(0)=a_0(n)>0$, meaning that $p_n$ must have a
negative real root $\lambda_1(n)<0$.  On the other hand, note that
\[
p'_n(\la)= 3\lambda^2+2a_2(n)\la+a_1(n)
\]
has roots given by
\[
\rho_+(n) = -\frac{a_2(n)}3
+ \frac{\sqrt{a_2^2(n) - 3a_1(n)}}{3}, \ \ \rho_-(n) = -\frac{a_2(n)}3
- \frac{\sqrt{a_2^2(n) - 3a_1(n)}}{3}.
\]
Moreover, a direct computation shows that
\[
a_2^2(n) - 3a_1(n)= \alpha^2b^2d^2\omega^4 n^4 +\alpha\omega^2(3(b+d)- 2(c\alpha+\ga) )n^2 + c^2\alpha^2-c\alpha \gamma + \gamma^2 >0,
\]
for example in the case $\omega^2 >\frac{2(a+c)}{b^2d^2}:= \omega_1^2(a, b, c, d)$, since for any $c$, $\alpha$, and $\gamma$ we have that
\[
c^2\alpha^2-c\alpha \gamma + \gamma^2>0.
\]
In other words, for $|\omega|>\omega_0$, we have  that $\rho_{\pm}(n)$ is real for any  $n\in \Z$. Hereafter, we choose appropriately $|\omega|>\omega_0$ and $a, b, c, d$ to guarantee that  $a_1(n)<0$ for $n\in \Z\setminus \{0\}$, meaning that $\rho_{+}(n)>0>\rho_{-}(n)$ for $n\in \Z\setminus \{0\}$. So, for $n\in \Z\setminus \{0\}$,  the polynomial $p_n$ has  either two positive real roots $\lambda_2(n), \lambda_3(n)$ or two complex conjugate root $\lambda_2(n), \lambda_3(n) \in \C\setminus i\R$ with  $\lambda_3(n)= \overline{\lambda_2(n)}$ and $\Re(\lambda_2(n))>0$.

Under previous assumptions, we are able to describe the form of the eigenvalues for $\widehat \A(n)$ for $n\ne 0$. In this case, we have that
\[
\beta_3(n) = -\beta_4(n) =\sqrt{\la_2(n)} \in \C\setminus i\R, \ \ \beta_5(n) = -\beta_6(n) =\sqrt{\la_3(n)} \in \C\setminus i\R,
\]
where ${\la_2(n)} \in \C\setminus i\R$ with $\Im({\la_2(n)}) \geq 0$.
A direct computation shows  that $\beta_m(n)$ for $1 \leq m \leq 6$ is a single eigenvalue with right eigenvector
\[
v_m(n) = \left(in, \beta_m(n), \beta^2_m(n), \beta^3_m(n),
{in Q(\bt_m(n), n)},{in \beta_m(n)Q(\bt_m(n), n) }\right)^t, \]
where
$$
Q(\bt_m(n), n) =\frac{\Lambda_m(n)}{\Theta_m(n)}, \ \  \Theta_m(n)={\beta^2_m(n)-(c\alpha+n^2)}, \ \ \ \Lambda_m(n)=\omega c\alpha( b\beta^2_m(n)-(bn^2+1)).
$$
It is also straightforward to show that
left eigenvector $z_m(n)$ is given by
\[
z_m(n)=Q(m,n) \left( \frac{\beta^3_m(n)}{in},
{\beta^2_m(n)}, \beta_m(n), 1, \frac{-in\omega\beta_m(n)
(\gamma-\alpha d) }{\Theta_m(n)},  \frac{-in\omega (\gamma-\alpha d)
}{\Theta_m(n)} \right),
\]
where $Q(m, n)$ is taken such that
\[z_m(n)v_l(n)=\delta_m^l.
\]
If we introduce the matrices $Z(n)$ and $V(n)$ given by
\[
Z(n)=\begin{pmatrix}z_1(n)\\ z_2(n)\\ z_3(n)\\ z_4(n)\\ z_5(n)\\ z_6(n)
\end{pmatrix},
\quad
V(n)=(v_1(n),v_2(n),v_3(n),v_4(n), v_5(n), v_6(n)),
\]
we have $Z(n)\cdot V(n)=I_6$ and
\[
Z(n)\widehat{A}(k)V(k)=
diag( \beta_1(n), \beta_2(n), \beta_3(n), \beta_4(n),\beta_5(n), \beta_6(n)), \ \ n\in \Z.
\]
Now, we observe that given any vector $\widehat U (n)\in \R^6$, we may write
\[
\widehat{U}(n)=V (n) \cdot Z(n)\widehat{U}(n) = V (n)U^{\#}(n), \ \ \mbox{where} \ \  U^{\#}(n) = Z(n)\widehat{U}(n) =\begin{pmatrix}U^{\#}_1(n)\\ U^{\#}_2(n)\\ U^{\#}_3(n)\\ U^{\#}_4(n)\\ U^{\#}_5(n)\\ U^{\#}_6(n)
\end{pmatrix}.
\]
Using this representation, we have for $U =\underset{n\in \Z}\sum \widehat{U}(n)e^{inx}$ that
\begin{equation} \label{fourier}
U =
\sum_{n\in\Z}\sum_{m=1}^6 v_m(n)\US_\jj(k) e^{inx},
\qquad
\mathcal AU=
\sum_{n\in\Z}\sum_{\jj=1}^6  \beta_\jj(n)v_\jj(n)\US_\jj(n)e^{inx}.
\end{equation}
We define the projections $\pi_0$ and $\pi_1$  by
\begin{align}\label{pidef0}
\pi_0 U & =
\sum_{ n \in \Z}\sum_{m=1}^2 v_m(n)\US_\jj(k) e^{inx}\\
\pi_1 U& =
\sum_{n \in \Z}\sum_{m=4}^6 v_m(n)\US_\jj(k) e^{inx}. \label{pidef1}
\end{align}
Moreover, the roots of the polynomial $p_n$ defined in (\ref{la}) are explicitly given by
\begin{align}
\la_1 &  =  -\frac13{a_2(n)} + (S(n) + T(n)) \label{la1} \\
\la_2 &  =  -\frac13{a_2(n)} - \frac12 (S(n) + T(n)) + \frac{i\sqrt 3}2 (S(n)- T(n))\label{la2}\\
\la_3 &  =  -\frac13{a_2(n)} - \frac12 (S(n) + T(n)) - \frac{i\sqrt 3}2 (S(n) - T(n))\label{la3}
\end{align}
where $S(n)$ and $R(n)$ are numbers defined as
\[
S(n) = \sqrt[3]{{R(n)} + \sqrt{D(n)}}, \ \  T(n) = \sqrt[3]{{R(n)} - \sqrt{D(n)}},
\]
where  the discriminant $D(n)$ of $p_n$ is defined as
$$
D(n) = Q^3(n)+ R^2(n).
$$
with  $Q$ and $R$  given by
\begin{align*}
Q(n)&= \frac{3a_1(n)-a^2_2(n)}{9} \\
R(n)&= \frac{9a_1(n)a_2(n)-27a_0(n)-2a^3_2(n) }{54}.
\end{align*}
We also have that $S(n)+T(n) \in  \R$  and that  $S(n)-T(n) \in  \R$ for $D(n)\geq 0$, and $S(n)+T(n) \in  \R$  and that  $S(n)-T(n) \in  i\R$ for $D(n)< 0$. Then, from this formulas we conclude for $|n|$ large enough  that
\begin{equation}
|\sqrt{D(n)}|\simeq O(n^6), \ \ |{R(n)}|\simeq O(n^6) \label{On2},
\end{equation}
implying that
\begin{equation}
|{T(n)}|\simeq O(n^2) \ \ |{S(n)}|\simeq O(n^2), \ \  \la_m(n)|\simeq O(n^2) \label{On},
\end{equation}
and so, for $| n |$ large enough and  $1 \leq m \leq 6$,   we have that
\begin{equation}
| \beta_m(n)|\simeq O(|n|) \label{btOn}.
\end{equation}
Using this fact it is not difficult to verify that in terms of the coefficient vectors $U^{\#}(n)$ we have the following equivalence of norms:
\begin{equation}\label{normeq}
\|U\|_H^2 \sim \sum_{n\in\Z} (1+n^2)^2|\US(n)|^2,
\qquad
\|U\|_X^2 \sim \sum_{n\in\Z} (1+n^2)^3|\US(n)|^2.
\end{equation}
From the equivalences in \eqref{normeq} it is evident that
$\pi_0$ and $\pi_1$ are bounded on $H$ and on $X$ with
$\pi_0+\pi_1=I$, and it is clear that $AX_j\subset H_j$
where $X_j=\pi_jX$ and $H_j=\pi_jH$ for $j=0,1$.
This yields the spectral decompositions
$H = H_0 \oplus H_1$ and $ X = X_0 \oplus X_1$.

\subsection{Linear dynamics analysis - Hypotheses \bf{(H0)} and \bf{(H1)}.} In this subsection, we want  to establish the conditions under which we are able to solve the uncouple system to the linear level. First we consider the center subspace $H_0$. We define the bounded
$C_0$-group $\{S_0(y)\}_{y\in\R}$ on $H_0$ with infinitesimal generator $\mathcal A_0= \mathcal A\mid_{X_0}$ by
\begin{equation}\label{S0def}
S_0(y)U= \underset{n \in \Z}\sum \sum_{\jj=1}^2
v_\jj(n)\US_\jj(n) e^{\beta_\jj(n)y}e^{inx}.
\end{equation}
It is straightforward to see that the group $(S_0(y))_{y\in \R}$ satisfies the hypothesis {\bf (H0)}.

Now, we want to verify the hypothesis {\bf (H1)}. To do this, we  determine the Green function to solve the problem in the hyperbolic subspace
$H_1$. Let us consider the inhomogeneous linear equation
\begin{equation}\label{h1eq}
\frac{d}{dy}U(y)=\A_1U(y)+G(y)
\end{equation}
where $\A_1=\A|_{X_1}$. We need to observe that
 $|\Re(\beta_m(n)|\ge\varepsilon>0$ for all $n \in \Z$ and $m = 3, 4, 5, 6$ when $n \ne 0$. In fact, assume that $\rho= \nu+ i \beta $ is a root of the polynomial $p_n$ with $\nu>0$ (a direct computation shows that $\nu \ne 0$). So, we have that $p_n$ is divided by the quadratic polynomial
\[
q(\lambda)= (\lambda-\rho)(\lambda- \overline{\rho})= z^2- 2\nu \lambda + |\rho^2|,
\]
which implies that the residue $r(\lambda)$ must be zero,  meaning  that
$$
a_1(n) -|\rho|^2 +a_2(n)+2\nu=0, \ \ a_0(n) -|\rho|^2(a_2(n)+2\nu)=0.
$$
So, if we assume that $a_1(n)+a_2(n)<0$ (which holds for large $n$), then we have that
$$
2\nu= -a_1(n)-a_2(n) +|\rho^2|\geq |\rho|^2.
$$
We note that condition $a_1(n)+a_2(n)<0$ holds whenever $\omega^2 >\frac{2(a+c)}{b+d}:= \omega_2^2(a, b, c, d)$.
Now, from the Cauchy lower bounds for roots of monic polynomials (see \cite{TDP-2013}), we have for any root $\lambda_m(n)$ of $p_n$ that
 \begin{equation} \label{boundroot}
 |\lambda_m(n)| \geq \frac{|a_0(n)|}{\max\{1, |a_0(n)|+|a_1(n)|, |a_0(n)|+|a_2(n)| \}} \geq \frac{|a_0(n)|}{1 +2|a_0(n)|+|a_1(n)|+|a_2(n)| \}},
 \end{equation}
which implies that there exists $\varepsilon >0$ such that for any $|n|\geq 1$, $|\lambda_m(n)|>\varepsilon^2$. In other words,  we have for $m=3, 5$ that,
\begin{equation}
\Re(\lambda_m(n)) \geq \frac{|\lambda_m(n)|^2}{2}\geq {\varepsilon^2}.
\end{equation}
From these facts, we conclude for $m=3, 5$ and $|n| \geq 1$ that
\begin{equation}
(\Re(\beta_m(n)))^2\geq  \Re(\beta^2_m(n)) \geq \frac{|\lambda_m(n)|^2}{2}\geq {\varepsilon^2} \ \  \Leftrightarrow \ \ \Re(\beta_m(n))\geq   {\varepsilon}.
\end{equation}
 So, for any $|n|\geq 1$ and $m=3, 4, 5, 6$, we have that $|\Re(\beta_m(n))|>{\varepsilon}$.

\smallskip

Hereafter we will assume $m = 3, 4, 5, 6$ and  $0\le\varrho<\varepsilon$. If we suppose $U\in C^1(\R,H_1)\cap C(\R,X_1)$ is a solution belonging to
$H_1^\varrho$ and $G\in C(\R,X_1)\cap H_1^\varrho$. Now,  using the Fourier series  expansion in $x$ and multiplying
by the matrix $Z(n)$ yields the differential equation
\begin{equation}\label{hseq}
\frac{d}{dy} \US_\jj(n,y)=\beta_\jj(n)\US_\jj(n,y)+\GS_\jj(n,y)
\end{equation}
The functions $\GS_\jj(n,\cdot)$ and $\US_\jj(n,\cdot)$ belong
to $\R^\varrho$ ($Y=\R$ in \eqref{Ybeta}). From the fact that $|\Re(\bt_\jj(n))|\ge\varepsilon>\varrho$, we conclude  necessarily that
\begin{equation}
\US_u(n, y)= \int_\infty^y e^{\bt_u(k)(y-\tau)} \GS_u(n,\tau)\,d\tau,
\qquad
\US_s(n, y)= \int_{-\infty}^y e^{\bt_s(n)(y-\tau)} \GS_s(n,\tau)\,d\tau.
\label{h1u34}
\end{equation}
As a direct consequence,   any solution of \eqref{h1eq} in $H_1^\varrho$ is unique. We will see that the formulas (\ref{h1u34}) together with the representation for $U=\pi_1U$ in (\ref{pidef1}) allow us to establish the existence of a solution in $H^1_\varrho$. To see this, we decompose  equation (\ref{h1eq}) using projections into the ``unstable" and ``stable" subspaces.
The projections for $U \in H$ are
\begin{align}
\pi_u U& =
\sum_{n \in \Z}\sum_{m=3, 5} v_m(n)\US_\jj(n) e^{inx}. \label{uproj}\\
\pi_s U& =
\sum_{n \in \Z}\sum_{m=4, 6} v_m(n)\US_\jj(n) e^{inx}. \label{sproj}
\end{align}
Clearly $\pi_u$ and $\pi_s$ are bounded on $H$ and $X$
and $\pi_u+\pi_s=\pi_1$. Now, we introduce a Green's
function operator $S(y)$ defined for nonzero $y \in\R$ by

\begin{equation}\label{sudef}
S(y)U= \left\{
\begin{array}{rl}
-\underset{n \in \Z}\sum  \underset{m=3, 5} \sum v_m(n)\US_\jj(n) e^{\bt_m(n)y} e^{inx},
  & y<0,  \\ \\
\underset{{n \in \Z}}\sum  \underset{m=4, 6}\sum v_m(n)\US_\jj(n) e^{\bt_m(n)y} e^{inx},
& y>0.
\end{array}
\right.
\end{equation}
By the definition of the norm in $H$ and $X$ (see \eqref{normeq}), we see directly for $Y = H$ or $ X$ that,
\[
\|S(y)U\|_H \sim \sum_{n\in\Z} (1+n^2)^2(e^{2\Re(\bt_3(n))y}|U_3(n)|^2 +e^{2\Re(\bt_5(n))y}|U_5(n)|^2) \leq e^{2\varepsilon}\|U\|_H^2, \ \ y<0.
\]
and also that
\[
\|S(y)U\|_H \sim \sum_{n\in\Z} (1+n^2)^3(e^{-2\Re(\bt_3(n))y}|U_3(n)|^2 +e^{-2\Re(\bt_5(n))y}|U_5(n)|^2) \leq e^{-2\varepsilon y}\|U\|_H^2, \ \ y>0.
\]
Moreover, we already have that
\[
\|S(y)U\|_Y  \leq e^{-2\varepsilon |y|}\|U\|_Y^2, \ \ \mbox{$Y=H$ or $Y=X$}.
\]
As a consequence of the Mean Value Theorem applied to the  $f(w)=e^{-w}$ with $w > 0$, we have that
\[
\left|\frac{e^{-w}-1}{-w}\right| \le 1,
\]
which implies  for $t>0$ that
\[
\sup_{\lambda\ge\varepsilon}\lambda^{-1}|e^{-\lambda y}-1| \le
t, \qquad
\sup_{\lambda\ge\varepsilon}\lambda e^{-\lambda y } =
\left\{\begin{array}{lr}
\varepsilon e^{-\varepsilon y }, & \ \varepsilon y \geq 1,\\
\frac1{ey}, & \ \varepsilon y \leq 1.
\end{array}
\right.
\]
Following the same type of calculation and using previous facts, one can easily verify that for
some constant $C$ (independent of $y$) we have the following norm bounds:

\begin{align}
\|S(y)\|_{\IL(Y)} &\leq C e^{-\varepsilon|y|}
\qquad\mbox{($Y=H$ or $X$),} \label{SHbd}\\
\|S(y)\|_{\IL(H,X)} &\leq
\left\{
\begin{array}{lr}
C e^{-\varepsilon y}, &  \ \varepsilon|t|\geq1,\\
C|y|^{-1},  &\ \varepsilon|y|\leq 1,
\label{SHXbd}
\end{array} \right.\\
\|S(y)-\pi_s\|_{\IL(X,H)}+\|S(-y)+\pi_u\|_{\IL(X,H)}&\leq  Cy, \ \ y>0,
\label{SXHbd}
\end{align}
where $\IL(Y)$ and $\IL(H,X)$ respectively denote the space of bounded operators on $Y$, and from $H$ to $X$. Clearly, we have that $S(y)\to\pi_s$ (resp. $-\pi_u$) strongly as $y\to0^+$ (resp. $0^-$).
Therefore the families $\{S(y)\}_{y>0}$ and $\{-S(-y)\}_{y>0}$ are analytic semigroups in $\pi_sH$ and $\pi_uH$ respectively \cite[p.62]{Pa}. Moreover, we also have that $S$ is $C^1$ from $\R\setminus\{0\}$ to $\IL(H)$ with $dS(y)/dy=A_1S(y)$. Therefore, we conclude that equation ~\eqref{h1u34}  yield the formula
\begin{equation}\label{h1u}
U(y)=\int_{-\infty}^\infty S(y-\tau)G(\tau)\,d\tau
\end{equation}
for the solution of \eqref{h1eq}. In order to establish that $U\in C(\R,X)$, $U\in H_1^\varrho$, and $dU/dy$ exists in $H$ and satisfies \eqref{h1eq}, we only need to follow the standard computations made in the case of  the Benney-Luke equation in \cite{PQ2}. In other words, we have verified the hypothesis {\bf (H1)}. So, we have established that system \eqref{eq:stw} admits a local center manifold having the properties stated in the Theorem (\ref{T:1}).

\section{\bf Global existence and stability for $b=d$, $a>d$ and  $|\omega|>1$ (large enough)}
We are now interested in proving global existence of classical
solutions on the local center manifold, for initial data that is
small in $H$-norm, which follows from the fact that the zero
solution is stable on the center manifold characterized by the graph
of a function $\phi_{\delta}: H_0 \to X_1$. We use strongly the
existence an energy functional that is conserved in time for
classical solutions in the case $b=d$ and $a>d$.

\nd {\bf Energy for the $abcd$-Boussinesq  system with $b=d$}.  Recall that we are using the variables
\[
u_1 = \partial_xv, \ \ \ \ u_2 =  \partial_yv,   \ \ \ \ u_3 =  \partial_{yy}v, \ \ \ \  u_4 = \partial_{yyy}v,  \ \ \ \ u_5 = u,   \ \ \ \ u_6 = \partial_yu,
\]
then we have  for $b=d$ that the travelling wave system takes the form

\begin{align}
-\omega u_x +d\omega u_{xxx}+d\omega u_{xyy}+v_{xx}+v_{yy}-cv_{xxxx}-2cv_{xxyy}-cv_{yyyy}+(uv_x^p)_x
+(uv_y^p)_y&=0, \label{s1}\\
\omega  v_x- d\omega v_{xxx}-d\omega v_{xyy}-u+a
u_{xx}+au_{yy}-\frac{1}{p+1}\left(v_x^{p+1}+v_y^{p+1}\right)&=0
\label{s2}.
\end{align}
Now, if we assume that $u, v$ are sufficiently smooth, then  a direct computation shows that
\begin{align}
\frac{1}{2\pi} \int_{0}^{2 \pi}\omega v_x u_y \,dx &=\frac{1}{2\pi} \int_{0}^{2 \pi}(\omega \partial_y(v_x u ) +\omega v_{y} u_x )\,dx\\
\frac{1}{2\pi} \int_{0}^{2 \pi} d\omega v_{xxx} u_y \,dx &=\frac{1}{2\pi} \int_{0}^{2 \pi}(d\omega \partial_y(v_{xxx} u ) +d\omega v_{y} u_{xxx} )\,dx\\
\frac{1}{2\pi} \int_{0}^{2 \pi} d\omega v_{xyy} u_y \,dx &=\frac{1}{2\pi} \int_{0}^{2 \pi}(d\omega \partial_y(v_{xy} u_y ) +d\omega v_{y} u_{xyy} )\,dx\\
\frac{1}{2\pi} \int_{0}^{2 \pi}((uv_x^p)_x +(uv_y^p)_y)v_y\,dx &= \frac{1}{2\pi} \int_{0}^{2 \pi}\left(\partial_y\left(u v_y^{p+1}\right)- \frac{u}{p+1}\partial_y\left(v_x^{p+1}+v_y^{p+1}\right)\right)\,dx\\
\frac{1}{2\pi} \int_{0}^{2 \pi} v_{yyyy}v_y \,dx&= \frac{1}{2\pi} \int_{0}^{2 \pi} \left(\partial_y(v_{yyy}v_y) -\frac12 \partial_y(v_{yy}^2) \right)\,dx
\end{align}
Multiplying equation (\ref{s1}) by $v_y$, equation (\ref{s2}) by $\partial_y u$,  and subtracting them, we have that
\begin{multline}
 \partial_y\left(\frac{1}{2\pi} \int_{0}^{2 \pi} \biggl( -|v_x|^2+
|v_y|^2 -c |v_{xx}|^2 +2c|v_{xy}|^2 +
c|v_{yy}|^2-|u|^2-a|\partial_xu|^2 +a|u_y|^2\right.\\ \left.
+2\omega v_x u -2d\omega v_{xxx}u-2d\omega v_{xy}u_y-2c v_{yyy}v_y+
\frac{2}{p+1} u \left(pv_y^{p+1} - v_x^{p+1}\right)\biggr)
\,dx\right) =0, \label{E0}
\end{multline}
So, in the variables $u_1, u_2, u_3, u_4, u_5, u_6$, we have that there exists an energy functional $\E\colon H \to {\R}$ defined by $\E(U)=\E_0(U)+ \E_1(U)$ which is conserved with respect to $y$ such that the quadratic part $\E_0(U)$ is given by 
\begin{multline}
\E_0(U)= \frac{1}{2\pi} \int_{0}^{2 \pi} \biggl( -|u_1|^2+
|u_2|^2 -c |\partial_x u_1|^2 +2c|\partial_x u_2|^2 + c|u_3|^2-|u_5|^2-a|\partial_x u_5|^2 +a|u_6|^2\\
+2\omega u_1 u_5 -2d\omega \partial_x u_2 u_6\biggr) \,dx -
\frac{d\omega}{\pi}\left( \partial_{xx}u_1, u_5\right)_{-1,1} -
\frac{c}{\pi}\left( u_4, u_2\right)_{-1,1} =0, \label{E0}
\end{multline}
 and the nonlinear part $\E_1$ is defined by
\begin{equation}\label{E1}
\E_1(U) = \frac{1}{\pi(p+1)}  \int_{0}^{2 \pi}u_5 (pu_2^{p+1} - u_1^{p+1}) \,dx,
\end{equation}
where $\left( \cdot, \cdot \right)_{-1,1}$ represents the pairing between  $\IH^{-1}$ and $\IH^{1}$.
From the definition, $\E$ is a smooth function from $H$ to $\R$. After multiplying appropriately the
equation (\ref{du4}) by $u_2$ and equation (\ref{du6}) by $u_6$, one can easily verify that
if $U \in C^1({\R},H)$ is a classical solution of the first order
equation \eqref{eq:stw}, then for all $y\in\R$
\[
\displaystyle{\frac{d}{dy}} \E(U(y)) = 0.
\]
Even though we have the conservation of the energy $\E$, we can not  use this  to obtain a solution throughout
the variational method since neither $\E$ nor $\E_0$ is not positive defined in the space $H$. However, we will see that energy $\E_0$ is positive on the center space $H_0$, and also that this controls the norm of $U$ in $H$, via the center manifold result. From the definition of the variable
$U = (u_1,  u_2, u_3, u_4, u_5, u_6)$,   we have a priori that  $u_1 = \partial_x v$ has mean zero on $[0, 2\pi]$, meaning that $\widehat U_1(0) = 0$.

\begin{lemma} \label{lm:E0}
Let $|\omega|\geq\max\{\omega_0, \omega_1, \omega_2\}$ and $a>d$. Then there is a positive constant $M_0>1$ such that for any $U \in H_0$ with $\widehat U_1(0) = 0$,
\begin{equation} \label{e0-norm}
{M_0^{-1}}\|U\|^2_{H} \leq \E_0(U) \leq M_0 \|U\|^2_{H}.
\end{equation}
\end{lemma}
\nd {\bf Proof.}
From the Fourier series representation of $ U \in H_0$ given with $\widehat U_1(0) = 0$, we have that
\begin{equation*}
U= \begin{pmatrix}0\\ \US_2(0)\\ 0\\0 \\ 0\\0\end{pmatrix}+ \underset{|n|>1} \sum \begin{pmatrix}in (\usp) \\ \bt_1(n)(\usm)\\
\bt_1^2(n)(\usp)\\ \bt_1^3(n)(\usm)\\{in Q_1(\bt_1(n), n)}(\usp) \\ {in\bt_1(n) Q_1(\bt_1(n), n)}(\usm)\end{pmatrix} e^{inx}= \widehat U(0)+  \underset{|n|>1} \sum \widehat U(n) e^{inx}.
\end{equation*}
We also have that $\E_0(U)= \underset{n\in \Z} \sum\E_0\left( \widehat U(n) e^{inx}\right)$. We note that
\[
\E_0(\hat U(0)) = |\US_2(0)|^2,
\]
On the other hand, for $\bt_1(n) = \bt$ we have that
\[
\E_0\left( \widehat U(n) e^{inx}\right)= \Ga_1(n)|\usp|^2+\Ga_2(n)|\usm|^2
\]
where
\begin{align*}
 \Ga_1(n)&= -n^2-cn^4+c\bt^4-n^2(1+an^2)Q^2(\beta,n) +2\omega n^2(1+dn^2)Q(\beta,n) \\
 \Ga_2(n)&=|\bt|^2\left( 1+n^2\left(2c+aQ^2(\beta,n)+2dwQ(\beta,n)\right) -2c\bt^2\right)
\end{align*}
Recall that we have that $\bt^2_1(n) = \la_1(n)
< 0 $  for $|n|>0$, so we conclude that  $Q(\beta,n)>0$ for $|n|>0$.
Thus, we have the right side of the second term $\Gamma_2$ is positive. Now,
for the first term note that
\[
-n^2-cn^4+c\bt^4-n^2(1+an^2)Q^2(\beta,n) +2\omega
n^2(1+dn^2)Q(\beta,n)= \frac{L_1+L_2}{\Theta_1^2(n)},
\]
where $L_1$ and $L_2$ are given by
\[
L_1 = \left(c\bt^4 - n^2(1+cn^2)\right)\Theta_1^2(n), \ \ L_2 =
2\omega n^2\left(1+dn^2\right)\Lambda_1(n)\Theta_1(n) -
n^2\Lambda_1^2(n)(1+an^2)
\]
Then we have that
\begin{align*}
L_1 +L_2&= c\bt^8+\vartheta_3(n)\bt^6 +
\vartheta_2(n)\bt^4+\vartheta_1(n)\bt^2 +\vartheta_0(n),
\end{align*}
with
\begin{align*}
\vartheta_3(n)&=-2c(c\alpha+n^2), \\
\vartheta_2(n)&=c(c\alpha+n^2)^2+\left[c(\omega^2\alpha d^2-1)n^2
+\omega^2c\alpha d(1-c\alpha d)+(\omega^2c\alpha d-1)\right]n^2,\\
\vartheta_1(n)&=-2\left[c(\omega^2\alpha
d^2-1)n^4+\left((\omega^2c\alpha d-1)+c\alpha(\omega^2d-c)\right)
+c\alpha(\omega^2-1)\right]n^2,\\
\vartheta_0(n)&=\left[c(\omega^2\alpha
d^2-1)n^4+\left((\omega^2c\alpha d-1)+c\alpha(\omega^2d-c)\right)
+c\alpha(\omega^2-1)\right](c\alpha+n^2)n^2.
\end{align*}
Note that $\vartheta_3(n)\leq0$. Now, using that
$\omega^2\geq\max\{1,\frac{c}{d},\frac{1}{\alpha bd}\frac{1}{c\alpha
b}\}$,   $\alpha ac=1$, $b=d$ and $a>d$, we see that
\[
\vartheta_2(n)\geq0,   \  \  \vartheta_1(n)\leq0, \   \
\vartheta_0(n)\geq0.
\]
Then using that $\bt^2_1(n) < 0$ for $|n|>0 $, we finally get that
\begin{equation*}
L_ 1 + L_2 \geq  0.
\end{equation*}
This fact implies that
\begin{multline*}
\underset{|n| >0}\min (\Ga_1(n), \Ga_2(n)) (|\usp|^2+|\usm|^2)\leq \E_0\left( \widehat U(n) e^{inx}\right)\\ \le \underset{|n| >0}\max (\Ga_1(n), \Ga_2(n)) (|\usp|^2+|\usm|^2),
\end{multline*}
which implies that
\begin{multline*}
\underset{|n| >0}\min (\Ga_1(n), \Ga_2(n)) (|\US_1(n)|^2+|\US_2(n)|^2)\\ \leq \E_0\left( \widehat U(n) e^{inx}\right)\le \underset{|n| >0}\max (\Ga_1(n), \Ga_2(n)) (|\US_1(n)|^2+|\US_2(n)|^2).
\end{multline*}
In other words, we have shown that
\[
\E(U)\sim \underset{ n \in \Z}\sum (1+n^2)^2\left(|\US_1(n)|^2+|\US_2(n)|^2\right)
\sim \|U\|_H^2.
\]
\qed

The first consequence of this fact is the following result:
\begin{corollary} \label{lm:E0}
Let $|\omega|\geq\max\{\omega_0, \omega_1, \omega_2\}$ and $a>d$.. Then there are $\delta_1>0$ and  $M_1>1$ such that for any $U \in H_0$ with $\widehat U_1(0) = 0$ and $\|U\|_{H}<\delta_1$ ,
\[
\frac1{M_1}\|U\|^2_{H} \leq |\E(U)| \leq M_1 \|U\|^2_{H}.
\]
\end{corollary}
{\bf Proof.} First note that from the Sobolev inequality, there is some positive constant $C = C(p)$ (independent of $U$) such that
\begin{align*}
|\E_1(U)| &=\left|\frac1{\pi(p+1)} \int_0^{2\pi} u_5 (pu_2^{p+1} - u_1^{p+1}) \,dx\right|\\
&\leq C \|u_5\|_{\tilde H^1}\left(\|u_2\|_{\tilde H^1}^{p+1}+\|u_1\|_{\tilde H^1}^{p+1}\right)\\
&\leq C \|U\|^{p+2}_{H}
\end{align*}
Moreover for some constants $C_0$ and $C_1 = C_1(p)$ (independent of $U$), we conclude that
\begin{align*}
|\E(U)| &\ge |\E_0(U)|- |\E_1(U)|\\
&\geq C_0 \|U\|^{2}_{H}- C_1 \|U\|^{p+2}_{H}\\
&\geq \|U\|^{2}_{H}\left(C_0 - C_ 1\|U\|^{p}_{H}\right).
\end{align*}
Let $\delta_1 > 0$ be such that $C_0-\delta_1^p C_1 > 0$. Then for $\|U\|_{H}\le \delta_1$ with $U\in H_0\setminus \{0\}$ we have that
\[
|\E(U)|\geq \|U\|^{2}_{H}\left(C_0 - C_1 \delta_1^p\right).
\]
The second claim of this lemma follows directly. In fact, for $U \in  H$ we have that
\begin{align*}
|\E(U)| \leq C \left(\|U\|^{2}_{H}+\|U\|^{p+2}_{H}\right)\\
\leq C\|U\|^{2}_{H}\left(1+ \|U\|^{p}_{H}\right).
\end{align*}
\qed

Now, we are interested in estimating the energy $\E$ on the center manifold. In other words, we want to
obtain a similar estimates for the lift of $\E$ to the center manifold $\md$.

\begin{lemma} \label{co:3.2.2}
Let $|\omega|\geq\max\{\omega_0, \omega_1, \omega_2\}$ and $a>d$, and let $\phd$ as in  Theorem~\ref{T:1}.
Then there exist constants $\delta_2>0$ and $C_2>1$ such that
for all $\xi\in H_0$ with $\|\xi\|_H<\delta_2$ we have
\[
\frac{1}{C_2}\|\xi\|_H^2 \le \E(\xi+\phd(\xi))
\le C_2\|\xi\|_H^2.
\]
\end{lemma}
{\bf Proof.} Let us define the functional $\tilde \E : H_0 \to  \R$ by
\[
\tilde \E(\xi) := \E(\xi + \phd(\xi)),
\]
where the function $\phd$ is defined in Theorem (\ref{T:1}). First note that $\|\phd(\xi)\|_H=o(\|\xi\|_H)$. Since $\E$ is smooth and $\E(0) = 0$, then $\E'(\xi) = O(\|\xi\|_{H})$. As a consequence of this fact and that $\E_1(U) = O\left(\|U\|^{p+2}_{H}\right)$, we have for $\xi \in H_0$ that
\begin{align*}
\E(\xi+\phd(\xi))
&=\E_0(\xi)+O(\|\xi\|_H\|\phd(\xi)\|_H) +\E_1(\xi+\phd(\xi))\\
&= \E_0(\xi)+o(\|\xi\|_H^2) .
\end{align*}
as $\|\xi\|_H\to0$. Then by the previous result, we get the conclusion.

\qed

We first establish  that solutions starting in the center manifold are appropriately bounded.

\begin{lemma}\label{bound}
Let $|\omega|\geq\max\{\omega_0, \omega_1, \omega_2\}$ and $a>d$, and let $\phd$ as in  Theorem~\ref{T:1}. If $\xi\in X_0$ be such that $\widehat \xi_1(0)=0$ and that $\|\xi\|_X\le \delta_2$. There exists a unique classical
solution $U(\xi, \cdot)$ for the full problem (without cutoff) \eqref{eq:stw} on $\R$   with initial condition $\pi_0 \circ U(\xi, 0)= \xi$ such that on any open interval $J$ containing  $0$,
\[
\|U_0(\xi, y)\|_H\le  C_2\|\xi\|_H \ \ \mbox{for any $y \in  J$}.
\]
\end{lemma}
{\bf Proof.}  We may assume that $\delta_2$ small enough such that $\delta_2<<\delta$. Let $\xi\in X_0$  be such that $\|\xi\|_X\le \delta_2$. Now, from Theorem~\ref{T:1}, there exists a unique continuous function $U$ form $\R$ to the local center manifold $\mathcal M_{\delta}$ such that $\pi_0(U(0))=\xi$, which turns out to be a classical solution of the equation (\ref{eq:stw})  on any open interval $J\subset \R$ containing $0$ such that $\|\pi_0(U(y))\|_H\leq \delta$ for any $y\in J$.  On the other hand, since $U$ is a classical solution and the energy $\mathcal E$ is conserved, then we have for any $y\in J$ that
\[
\frac{1}{C_2}\||\pi_0(U(y))\|_H^2 \le \E(\pi_0(U(y)))= \mathcal E(U(0))
\le C_2\|\xi\|_H^2,
\]
meaning  $||\pi_0(U(y))\|_H \le  C_2\|\xi\|_H  $, for any $y\in J$ as desired. A continuation argument shows that $U$ is a classical  solution for the full problem (without cutoff) \eqref{eq:stw} on $\R$.

\qed

Now we are in position to state the main result on the existence and the stability on the
center manifold. The proof of this result follows in the same fashion as the result for the Benney-Luke equation done by Quintero and Pego in \cite{PQ2}.

\begin{theorem}\label{th:stab}
{\bf (Global Existence and stability on the center manifold)}
Let $|\omega|\geq\max\{\omega_0, \omega_1, \omega_2\}$ and $a>d$, and let $\phd$ be given by applying Theorem~\ref{T:1} to \eqref{eq:stw}.
There exist positive constants $\delta_3$ and $C_3$ such that,
for any $\xi\in X_0$ with $\widehat \xi_1(0)=0$ and  $\|\xi\|_H\le \delta_3$,
there is a unique classical solution $U$ on $\R$ to \eqref{eq:stw} such that
$\pi_0 U(0)=\xi$ and
$\|U(y)\|_H\le 2C_2\|\xi\|_H$ for all $y\in\R$.
Moreover, for any $T>0$ the map taking $\xi$ to $U$ is Lipschitz continuous
from $H_0$ to $C([-T,T],H)$.
\end{theorem}

\if{
{\bf Proof}
Let $\delta_2$ and $C_2$ be given by Lemma \ref{0-est}.
We may assume $2C_2\delta_2<\delta$ by making $\delta_2$ smaller
if necessary.
Suppose $\xi\in X_0$ with $\|\xi\|_H\le\delta_2$.
We invoke Theorem~\ref{T:1} and obtain existence of a continuous function
$U$ from $\R$ into the center manifold
$\md=\{\zeta+\phd(\zeta):\zeta\in X_0\}$ such that $\pi_0U(0)=\xi$,
which is a classical solution of \eqref{eq:stw} on any
open interval $J\subset\R$ containing $0$ such that
$\|\pi_0U(y)\|_H<\delta$ for all $y\in J$.

By a straightforward continuation argument, to show that $U$
is a classical solution on $\R$ and satisfies the estimate
claimed in Theorem \ref{th:stab}, it suffices to establish
an appropriate a priori estimate. Namely, it is enough
to show that on any
open interval $J\subset\R$ containing $0$ such that
$\|\pi_0U(y)\|_H<\delta$ for all $y\in J$,
we have that
$\|\pi_0U(y)\|_H\le C_2\|\xi\|_H$ for all $y\in J$.

Since $U$ is a classical solution on such an interval, we
may use the fact that $\E(U(y))$ is constant along with
Lemma \ref{0-est} to deduce that for any $y\in J$,
\begin{equation}\label{4-eest}
\frac{1}{C_2}\|\pi_0U(y)\|_H^2 \le \E(U(y)) =\E(U(0))
\le C_2\|\xi\|_H^2.
\end{equation}
This establishes the desired a priori estimate, proving
the existence part of the Theorem. The statement regarding
Lipschitz dependence follows from Proposition \ref{R:2.4}.

To prove the uniqueness statement, suppose
$U$ is a classical solution on $\R$ to \eqref{eq:stw} such that
$\pi_0 U(0)=\xi$ and
$\|U(y)\|_H\le 2C_2\|\xi\|_H$ for all $y\in\R$.
Since $2C_2\|\xi\|_H<\delta$,
from Theorem~\ref{T:1}(v) it follows $U(y)$ must lie on the center manifold
$\md$ for all $y\in\R$.
Then $U$ is determined by $U_0=\pi_0U$, which by Theorem~\ref{T:1}(iii)
is the unique classical solution of the equation
\begin{equation}
\label{eq:U0}
 \frac{d}{dy}U_0(y) = A_0U_0(y) + \pi_0 f(U_0(y)+\phd(U_0(y)))
\end{equation}
with $U_0(0)=\xi$.

XXXXXXXXXXXXXXXXXXXXXXXXXXXXx


\subsection{\bf On the parametrization and the evolution on the center manifold.}
In this section,
we will see that the solution on the center manifold can be characterized only by their initial data
of the first two components. This will be establish by looking at the correspondence between $U(0)$ and $\pi_0U(0)$. For notational convenience, we define the spaces $\V_j$ and $\W_j$ for $j = 1; 2$
\begin{align*}
\V_1& =\left\{U\in H: \sum_{0\leq \mid n\mid \leq n_0}\left(\sum_{m=1}^2 v_m(n)\US_\jj(k) \right)e^{inx} \right\}\\
\W_1& =\left\{U\in H: \sum_{0\leq \mid n\mid \leq n_0}\left(\sum_{m=3}^6 v_m(n)\US_\jj(k)\right) e^{inx}+\sum_{ \mid n\mid > n_0}\left(\sum_{m=1}^6 v_m(n)\US_\jj(k) \right)e^{inx} \right\}\\
\V_1& =\left\{U\in X: \sum_{0\leq \mid n\mid \leq n_0}\left(\sum_{m=1}^2 v_m(n)\US_\jj(k) \right)e^{inx} \right\}\\
\V_1& =\left\{U\in X: \sum_{0\leq \mid n\mid \leq n_0}\left(\sum_{m=3}^6 v_m(n)\US_\jj(k) \right)e^{inx}+\sum_{ \mid n\mid > n_0}\left(\sum_{m=1}^6 v_m(n)\US_\jj(k) \right)e^{inx} \right\}
\end{align*}
In other words, $\V_1 = H_0$, $\V_2 = X_0$, $\W_1 = H_1$ and $\W_2 = X_1$. Now, let us denote with $\mathcal R$ the restriction operator of any $U = (U_1, U_2, U_3, U_4, U_5, U_6)^t \in H$ (or $X$) on the first two components by
\[
\Ronetwo U = \begin{pmatrix}U_1\cr U_2\end{pmatrix}=\sum_{n\in\Z}
\vonetwo(n, \be_1(n))\begin{pmatrix}\US_1(n)\cr\US_2(n)\end{pmatrix} e^{inx}.
\]
where the $2\times 2$ matrix $\vonetwo(n, \bt)$ is defined as
\[
\vonetwo(0, \be)=
\begin{pmatrix}
1&0 \\ 0&1
\end{pmatrix}, \quad
\vonetwo(n, \bt)=
\begin{pmatrix}in & in\\ \bt&-\bt
\end{pmatrix}
\]
From the definition, it is clear that if $U \in \V_j$ for $j = 1, 2$, then we have that
\[
\Ronetwo U = \begin{pmatrix}U_1\cr U_2\end{pmatrix}=\underset{0\leq \mid n\mid \leq n_0}\sum
\vonetwo(n, \bt_1(n))\begin{pmatrix}\US_1(n)\cr\US_2(n)\end{pmatrix} e^{inx}.
\]
Clearly, $\mathcal R$ is bounded from $H_0$ to $K^j_{n_0} = \pi^{n_0}(\tilde H^j \times \tilde H^j)$, where $\pi^{n_0}$ is the projection of $\tilde H^j \times \tilde H^j$ in the first $|n_0|$ components. On the other hand, since for all $n$ the matrix $\vonetwo(n, \bt_1(n))$ is
invertible, then $\mathcal RU = 0$ implies that $\mathcal RU(n) = 0$. This means that $\mathcal R$ is one-to-one. We will see in the coming result that the inverse of $\mathcal R$ is given by the prolongation formula $U = \mathcal P(w) \in  H_0$ for given $w = (w_1, w_2)^t \in K^j_{n_0}$ with $j = 1, 2$.

We summarized the properties of the operators $\mathcal R$ and  $\mathcal P$ in the following result.
\begin{lemma}
For $j = 1, 2$, the application  $\mathcal R$ is an isomorphism from the space $\V_j$  to $K^j_{n_0}$. In other words, $H_0 \cong K^1_{n_0}$ and $X_0 \cong K^2_{n_0}$.
\end{lemma}
{\bf Proof.} Let $U\in \V_j$. From the representation of $U$, we see directly
\[
\Ronetwo U=\underset{0\leq \mid n\mid \leq n_0}\sum
\vonetwo(n, \bt_1(n))\begin{pmatrix}\US_1(n)\cr\US_2(n)\end{pmatrix} e^{inx}.
\]
A direct computation shows for $0\leq \mid n\mid \leq n_0 $ that
\begin{align*}
|\widehat{\Ra U}(n)|^2&\leq C\left(|in \US_1(n)+ in \US_2(n)|^2 + |\bt_1(n) \US_1(n)- \bt_1(n)  \US_2(n)|^2\right)\\
&\leq C\left(n^2(| \US_1(n)|^2 + | \US_2(n)|^2) +n^2 (|\bt_1(n) \US_1(n)|^2+  |\US_2(n)|^2)\right)\\
&\leq C(1+n^2)\left(| \US_1(n)|^2 + | \US_2(n)|^2\right)
\end{align*}
This implies that $\Ra$ is a bonded linear operator from $\V_j$  to $K^j_{n_0}$ for $j = 1, 2$. It is also straightforward to verify that $\Ra$ is one to one on $\V_j$ and $\W_j$ for $j = 1, 2$. On the other hand, given $w = (w_1, w_2)^t \in K^j_{n_0}$  for $j = 1,  2$, we can define for $\US_1(n)$ and $\US_2(n)$ for each $0\leq \mid n\mid \leq n_0 $ as
\begin{align}
\begin{pmatrix}\US_1(0)\cr\US_2(0)\end{pmatrix} &= \begin{pmatrix}\hat w_1(0)\cr\hat w_2(0)\end{pmatrix} \\
\begin{pmatrix}\US_1(n)\cr\US_2(n)\end{pmatrix}&=  \vonetwo^{-1}(n, \bt_1(n)) \hat w_(n)= \frac{1}{2in\bt_1(n)}\begin{pmatrix}\bt_1(n) & in\\\bt_1(n) & -in
\end{pmatrix}\begin{pmatrix}\hat w_1(n)\cr\hat w_2(n)\end{pmatrix}
\end{align}
This facts let us define the extension operator $Pa$ on the space $ K^j_{n_0}$ for $j = 1,  2$. In fact, if $U = \Pa(w)$, then we set $U$ in $H_0$ or $X_0$ as as follows
\begin{equation}\label{inversa_R}
U=\begin{pmatrix} \US_1(0)\\ \US_2(0)\\ 0\\0 \\ 0\\0\end{pmatrix}+ \underset{0\leq |n| \leq n_0} \sum \begin{pmatrix}in (\usp) \\ \bt_1(n)(\usm)\\
\bt_1^2(n)(\usp)\\ \bt_1^3(n)(\usm)\\- \frac{cn\ga i}{\Theta_1(n)}(\usp) \\ -\frac{cn\ga \bt_1(n) i}{\Theta_1(n)}(\usm)\end{pmatrix} e^{inx}.
\end{equation}Then it is clear for $0\leq |n| \leq n_0$ that
\[
| \US_1(n)|^2 + | \US_2(n)|^2
\]

\begin{theorem} \label{th:init}
Let $\phd$ be given by applying Theorem~\ref{T:1} to \eqref{eq:stw}.
There exist positive constants $\delta_3$ and $C_3$ with the
following property. For any $w=(w_1,w_2)\in\HHone$ such that
$\|w\|_{\HHone}<\delta_3$, there exists a unique
$\xi\in H_0$ such that $\|\xi\|_H\le C_3\|w\|_\HHone$ and
\begin{equation}\label{5-w}
w=\Ronetwo (\xi+\phd(\xi)).
\end{equation}
The map $w\to\xi$ is Lipschitz continuous, and if $w\in\HHtwo$ then
$\xi\in X_0$.
\end{theorem}

To prove this, first we study the restriction $\Ronetwo$ on the center subspace.

\begin{lemma} \label{lm:r12}
The map $\Ronetwo$ yields simultaneous isomorphisms
$H_0 \cong \HHone$ and $X_0\cong\HHtwo$.
\end{lemma}

\proof
Given any $U=\pi_0 U$ in $H_0$ or $X_0$, we use the representation
\eqref{pidef} to write
\begin{equation}\label{5-ru}
\Ronetwo U = \sum_{k\in\Z} e^{ikx}
\vonetwo(k)\begin{pmatrix}\US_1(k)\cr\US_2(k)\end{pmatrix}.
\end{equation}
where
\begin{equation}\label{5-v12}
\vonetwo(0)=
\begin{pmatrix}
1&0 \\ 0&1
\end{pmatrix}\quad
\vonetwo(k)=
\begin{pmatrix}ik & ik\\ \lambda_1(k)&-\lambda_1(k)
\end{pmatrix}
\end{equation}
for $k\ne0$. Since $|\lambda_1(k)|$ grows asymptotically linearly in $k$,
it follows that
\begin{equation}\label{5-rest}
|\widehat{\Ronetwo U}(k)|^2 \le C(1+k^2)(|\US_1(k)|^2+|\US_2(k)|^2).
\end{equation}
From the equivalences \eqref{normeq} it follows that
$\Ronetwo$ is bounded from $H_0$ to $\HHone$ and from $X_0$ to $\HHtwo$.
Moreover, $\Ronetwo$ is one-to-one, since if $\Ronetwo U=0$ then
$\widehat{U}(k)=0$ for all $k$ since $\vonetwo(k)$ is invertible.

To see that $\Ronetwo$ yields an isomorphism, we observe that its inverse
is given by the prolongation formula
$U=\Ronetwoinv w$, where, given $w\in \HHone$
or $\HHtwo$, we have $U=\pi_0U$ given by \eqref{pidef} with

\begin{equation}\label{5-rinv}
\begin{pmatrix}\US_1(k)\\ \US_2(k)\end{pmatrix}
= \vonetwo(k)^{-1}\hat{w}(k) =
\frac{1}{2ik\lambda_1(k)}
\begin{pmatrix}\lambda_1(k)&ik\\\lambda_1(k)&-ik\end{pmatrix}
\begin{pmatrix}\hat{w}_1(k)\\\hat{w}_2(k)
\end{pmatrix}
\end{equation}

for $k\ne0$. Clearly
\[
|\US_1(k)|^2+|\US_2(k)|^2=
|\vonetwo(k)^{-1}\hat{w}(k)|^2 \le \frac{C}{1+k^2}|\hat{w}(k)|^2
\]
for all $k$, so using \eqref{normeq} we see that
$\Ronetwoinv$ is bounded from $\HHone$ to $H_0$ and from
$\HHtwo$ to $X_0$.
\qed

\noindent
{\bf Proof of Theorem \ref{th:init}.}
Given $w$, we shall prove corresponding statements for
$\zeta=\Ronetwo\xi$ in place of $\xi$ and use Lemma~\ref{lm:r12}.
Recall that $\Ronetwoinv$, the inverse of $\Ronetwo$, is given by
$U=\Ronetwoinv w$ using \eqref{5-rinv} and \eqref{pidef}.
The quantity $\zeta$ should satisfy
\begin{equation}\label{5-zeta}
\zeta = w-\ponetwo(\zeta)\qquad\hbox{where}\quad
\ponetwo(\zeta)=\Ronetwo \phd(\Ronetwoinv\zeta).
\end{equation}
Using the contraction mapping theorem, we find that this
equation has a unique solution $\zeta\in\HHone$ satisfying
$\|\zeta\|_\HHone<\delta'$ if $\|w\|_\HHone<\delta'(1-L(\delta'))$,
provided that the Lipschitz constant $L(\delta')$ of $\ponetwo$
on the ball of radius $\delta'$ in $\HHone$ satisfies $L(\delta')<1$.
This is true if $\delta'$ is sufficiently small, due to Theorem~\ref{T:1}(ii)
and Lemma~\ref{lm:r12}.
Moreover, the map $w\mapsto\zeta$ is Lipschitz with Lipschitz constant
$1/(1-L(\delta'))$. Furthermore, $\ponetwo(\zeta)\in\Ronetwo
X_1\subset \HHtwo$, so if $w\in\HHtwo$ then $\zeta\in\HHtwo$ and so
$\xi=\Ronetwoinv\zeta\in X_0$.
\qed

\section*{Acknowledgments}
This material is based upon work supported by the
National Science Foundation under Grant
Nos.\  DMS97-04924 and DMS00-72609.
The work of JRQ is supported by the
Universidad del Valle, Cali, Colombia, and
partially supported by Colciencias under Project No.\ 1106-05-10097.

\noindent
{\bf Proof of Theorem~\ref{th:main}.}
The parts of the Theorem referring to existence, uniqueness,
stability, and Lipschitz dependence on initial data follow directly from
Theorems~\ref{th:stab} and \ref{th:init}. It remains to verify that
the first two components $w=\Ronetwo U$ of a solution as given by
these results satisfy an equation of the form \eqref{eq:wave}.
Let $\xi(y)$ be determined from $w(y)$ by Theorem~\ref{th:init} for
each $y$. Since $\Ronetwoinv\Ronetwo\xi=\xi$, we find that
\begin{equation}\label{eq:uw}
U = \Ronetwoinv w +\wtoU(w) \qquad\hbox{where}\quad
\wtoU(w) = (I-\Ronetwoinv\Ronetwo)\phd(\xi).
\end{equation}
Note that $\phd$ takes values in $X_1$ so $\wtoU(w)$ need not
be zero.  However the first two components $\Ronetwo\wtoU(w)=0$,
and since $U$ is a classical solution of \eqref{eq:stw}, restriction
to the first two components yields
\begin{equation}\label{eq:w1}
\frac{d}{dy} w= \Ronetwo A (\Ronetwoinv w+\psi(w)) = \Aonetwo w +
\begin{pmatrix}  0\cr \psi_3(w)\end{pmatrix}
\end{equation}
since the first two components $\Ronetwo f(U)=0$.
Here the action of the operator $\Aonetwo=\Ronetwo A\Ronetwoinv$
can be determined through Fourier transform representation from
\eqref{3-urep} and \eqref{5-rinv}. We find that
\begin{eqnarray}
\Aonetwo w &=& \sum_{k\in\Z} e^{ikx} \vonetwo(k)
\begin{pmatrix}
\lambda_1(k)&0\cr0&-\lambda_1(k)
\end{pmatrix}
\vonetwo(k)^{-1} \hat w(k)
\nonumber \\
&=& \sum_{k\ne0} e^{ikx}
\begin{pmatrix}0 & ik\cr \lambda_1(k)^2/ik &0\end{pmatrix}
\begin{pmatrix}\hat w_1(k)\cr \hat w_2(k)\end{pmatrix} .
\label{5-Arep}
\end{eqnarray}
Thus we have that $w=\Ronetwo U$ satisfies an equation of the form
\eqref{eq:wave} in which $g=\wtoU_3$ and
$S$ is a pseudodifferential operator of degree zero defined by
\begin{equation}
\label{eq:Shat}
\widehat{Sw_1}(k)=
(\lambda_1(k)/ik)^2 \hat w_1(k)
\end{equation}
for $k\ne0$.

The eigenvalues of $\Aonetwo$ have the form $\pm\lambda_1(k)$,
leading to the real dispersion relation in \eqref{1-lsq}
with the plus sign.
Since $\phd$ takes values in $X_1$ and satisfies $\phd(0)=0$,
$D\phd(0)=0$, we find that $g(w_1,w_2)=\psi_3(w)$ is Lipschitz from
a small ball in $\HHone$ into $\IH_1$ with $g(0,0)=0$, $Dg(0,0)=0$.

This finishes the proof of Theorem \ref{th:main}. \qed
}\fi

\section{\bf Parametrization and evolution on the center manifold}

In this section we assume that $|\omega|\geq\max\{\omega_0, \omega_1, \omega_2\}$ and $a>d$. We will see that any global solutions $U$ on the center manifold is  characterized by their initial data in the first two components. This can be accomplished by showing an appropriate correspondence between the first two components
and the center-subspace projection $\pi_0U(0)$. Hereafter, we denote the restriction of any $U=(u_1,u_2,u_3,u_4, u_5, u_6)^T$ on the first two components by
\be\label{5-r12}
\Ronetwo U = (u_1,u_2).
\ee

\begin{lemma} \label{lm:r12}
The map $\Ronetwo$ yields simultaneous isomorphisms
$H_0 \cong \HHone$ and $X_0\cong\HHtwo$.
\end{lemma}

\proof
Given any $U=\pi_0 U$ in $H_0$ or $X_0$, we use the representation
\eqref{pidef0} to write
\be\label{5-ru}
\Ronetwo U = \sum_{n\in\Z}
\vonetwo(n)\begin{pmatrix}\US_1(n)\\\US_2(n)\end{pmatrix}e^{inx}.
\ee
where
\be\label{5-v12}
\vonetwo(0)=\begin{pmatrix}1&0\\ 0&1\end{pmatrix},\quad
\vonetwo(n)=
\begin{pmatrix}in & in  \\ \lambda_1(n)&-\lambda_1(n)\end{pmatrix}
\ee
for $n\ne0$. Since $|\lambda_1(n)|$ grows asymptotically linearly in $n$,
it follows that
\begin{equation}\label{5-rest}
|\widehat{\Ronetwo U}(n)|^2 \le C(1+n^2)(|\US_1(n)|^2+|\US_2(n)|^2+|\US_5(n)|^2).
\end{equation}
From the equivalences \eqref{normeq} it follows that
$\Ronetwo$ is bounded from $H_0$ to $\HHone$ and from $X_0$ to $\HHtwo$.
Moreover, $\Ronetwo$ is one-to-one, since if $\Ronetwo U=0$ then
$\widehat{U}(n)=0$ for all $n$ since the matrix $\vonetwo(n)$ is invertible.

To see that $\Ronetwo$ yields an isomorphism, we observe that its inverse
is given by the prolongation formula $U=\Ronetwoinv w$. More concretely, given $w\in \HHone$ or $\HHtwo$, we define $\US_1(n)$ and $\US_2(n)$ by
\be\label{5-rinv}
\begin{pmatrix}\US_1(n)\cr \US_2(n)\end{pmatrix}= \vonetwo(n)^{-1}\hat{w}(n) =
\frac{1}{2in\lambda_1(n)}
\begin{pmatrix}\lambda_1(n)&in\cr\lambda_1(n)&-in\end{pmatrix}
\begin{pmatrix}\hat{w}_1(n)\cr\hat{w}_2(n)\end{pmatrix}.
\ee
Then we define $\Ronetwoinv w$ as
\begin{equation*}
\Ronetwoinv w=
\sum_{ n \in \Z}\sum_{m=1}^2 v_m(n)\US_\jj(n) e^{inx}.
\end{equation*}
So, we have that $\Ronetwoinv$ is the inverse of $\Ronetwo$, and also using \eqref{normeq} we see that $\Ronetwoinv$ is bounded from $\HHone$ to $H_0$ and from
$\HHtwo$ to $X_0$, since for $n\ne0$. we have that \[
|\US_1(n)|^2+|\US_2(n)|^2=
|\vonetwo(n)^{-1}\hat{w}(n)|^2 \le \frac{C}{1+n^2}|\hat{w}(n)|^2
\]
We also have that $U=\pi_0U$ and that $\Ronetwo o \Ronetwoinv =I_{W}$, where $W$ is either $\HHone$ or $\HHtwo$, and also that $\Ronetwoinv o \Ronetwo   =I_{W}$, where $W$ is either $H_0$ or $X_0$.
\qed

\begin{theorem} \label{th:init}
Let $\phd$ be given by applying Theorem~\ref{T:1} to \eqref{eq:stw}.
There exist positive constants $\delta_3$ and $C_3$ with the
following property. For any $w=(w_1,w_2)\in\HHone$ such that
$\|w\|_{\HHone}<\delta_3$, there exists a unique
$\xi\in H_0$ such that $\|\xi\|_H\le C_3\|w\|_\HHone$ and
\be\label{5-w}
w=\Ronetwo (\xi+\phd(\xi)).
\ee
The map $w\to\xi$ is Lipschitz continuous, and if $w\in\HHtwo$ then
$\xi\in X_0$.
\end{theorem}
{\bf Proof}.
Let $w$ be given, then using that $\mathcal R$ is invertible, we see that equation (\ref{5-w}) is equivalent to solve for $\zeta= \mathcal R (\xi) \in\HHone $ the equation
\be\label{5-zeta}
w=\zeta +\Ronetwo (\phd( \Ronetwoinv(\zeta)))= \zeta +\ponetwo(\zeta) \ \ \Leftrightarrow \ \ \zeta = w-\ponetwo(\zeta),
\ee
$\ponetwo=\Ronetwo  \circ \phd \circ  \Ronetwoinv $.  We note from Theorem~\ref{T:1}(ii) that
\[
\|\ponetwo\|_{\HHone}\leq M || \phd(\Ronetwoinv\zeta)||_{\HHone}\leq M_1 L(\delta') || \zeta||_{\HHone},
\]
and also that $L(\delta')\to 0$,  if $\delta' \to 0$. So we conclude form the Contraction Mapping Theorem that equation (\ref{5-zeta}) has a unique solution $\zeta\in\HHone$ satisfying
$\|\zeta\|_\HHone<\delta'$ if $\|w\|_\HHone<\delta'(1-L(\delta'))$,
provided that the Lipschitz constant $L(\delta')$ of $\ponetwo$
on the ball of radius $\delta'$ in $\HHone$ satisfies $L(\delta')<1$. On the other hand, we also know that the map $w\mapsto\zeta$ is Lipschitz with Lipschitz constant
$1/(1-L(\delta'))$. Furthermore, $\ponetwo(\zeta)\in\Ronetwo
X_1\subset \HHtwo$, so if $w\in\HHtwo$ then $\zeta\in\HHtwo$ and so
$\xi=\Ronetwoinv\zeta\in X_0$.
\qed
\medskip

Now we are in position to establish the main result related with the dynamics on the center manifold.
\begin{theorem}(Traveling wave solutions via dynamics in $y$) \label{th:main}
There are positive constants $\delta_1$ and $C_1$ with the following
property:
Given any initial conditions of the form
\[
(u_1(0), u_2(0))= (w_1, w_2)
\]
in $\IH^2\times \IH^2$
such that 
$\left\|(w_1, w_2) \right\|_{ \IH^1\times\IH^1} \leq \delta_1$,
equation ~(\ref{eq:stw}) has a unique
global classical solution $U\in C^1(\R,H)\cap C(\R, X)$
such that
$\|U(y)\|_{H} \leq C_1\delta_1$ for all $y\in\R$.
The map taking initial conditions
to the solution is Lipschitz continuous
from $\IH^1\times\IH^1$ to $C([-T,T],H)$,
for any $T>0$.

Moreover, the first two components of the solution satisfy
a dispersive, nonlinear, nonlocal wave equation of the form
\be\label{eq:wave}
\frac{d}{dy}\begin{pmatrix}u_1\cr u_2\end{pmatrix} =
\begin{pmatrix}0&\dx\cr S\dx&0\end{pmatrix}\begin{pmatrix}u_1\cr u_2\end{pmatrix}+
\begin{pmatrix}0\cr g(u_1,u_2)\end{pmatrix},
\ee
in which the map
$g\colon \IH^1\times\IH^1\to \IH^1$ is Lipschitz with
$g(0)=0$, $Dg(0,0)=0$, and where the nonlocal
linear operator $S$ defined by equation \eqref{eq:Shat}.
\end{theorem}
{\bf Proof.}
The parts of the Theorem referring to existence, uniqueness,
stability, and Lipschitz dependence on initial data follow directly from
Theorems~\ref{th:stab} and \ref{th:init}. It remains to verify that
the first two components $w=\Ronetwo U$ of a solution as given by
these results satisfy an equation of the form \eqref{eq:wave}. For each $y\in \R$, we set  $\xi(y)$ to be determined from $w(y)$ by Theorem~\ref{th:init}, meaning that
\[
w(y)=\Ronetwo (\xi(y))+ \Ronetwo (\phd(\xi(y))), \ \ y\in \R.
\]
On the other hand, the projection $\pi_0$ on the center space $H_0$ commutes with restriction $\mathcal R$, implying that
\[
\pi_0 w(y)= \Ronetwo (\xi(y)),
\]
but we also have that
\[
\pi_0 w(y)= \pi_0(U(y))= \pi_0(U_0(y) + \phd(U_0(y))= \pi_0(U_0(y)),
\]
where $U_0$ is the solution on the center space. From this facts, we conclude that
$\xi(y) = U_0(y)$. Then using that $\Ronetwoinv\Ronetwo\xi=\xi$ and setting $\varphi(w) = (I-\Ronetwoinv\Ronetwo)\phd(\xi) $, we have that
\begin{align}
U(y)& = \xi(y) + \phd(\xi(y)) \nonumber \\
&=  \xi(y) +  \Ronetwoinv\Ronetwo(\phd(\xi(y)) + (I- \Ronetwoinv\Ronetwo)(\phd(\xi(y)) \nonumber  \\
&= \Ronetwoinv\left(\Ronetwo\left( \xi(y) + \phd(\xi(y))\right) \right)+ (I- \Ronetwoinv\Ronetwo)(\phd(\xi(y))\nonumber  \\
&= \Ronetwoinv w(y) +\varphi (w(y)). \label{eq:uw}
\end{align}
Now, we need to recall that $\phd$ maps $X_0$  into $X_1$, so $\varphi(w)$ need not
be zero, however the first two components $\Ronetwo \varphi(w)$ are zero. On the other hand,  $U$ is a classical solution of \eqref{eq:stw}, then we have that
\be\label{eq:w1}
\frac{d}{dy} w= \Ronetwo \left( AU+ G(U)\right)=  A (\Ronetwoinv w+\varphi(w)) = \Aonetwo w +
\begin{pmatrix} 0\cr \varphi_3(w)\end{pmatrix}
\ee
since we have that the first two components $\Ronetwo G(U)=0$ and that $\varphi(w)$ has the first two components zero. We only need to determine the action of the operator $\Aonetwo=\Ronetwo A\Ronetwoinv$. To do this, we use the Fourier transform representation from
\eqref{fourier} and \eqref{5-rinv}. We find that
\begin{eqnarray}
\Aonetwo w &=& \sum_{n\in\Z}
\begin{pmatrix} in \bt_1(n)&-in \bt_1(n)\cr \bt^2_1(n)& \bt^2_1(n)\end{pmatrix} \begin{pmatrix}\US_1(n)\cr \US_2(n)\end{pmatrix}e^{inx}
\nonumber \\
&=& \sum_{n\in\Z} \begin{pmatrix} in \bt_1(n)&-in \bt_1(n)\cr \bt^2_1(n)& \bt^2_1(n)\end{pmatrix}  \vonetwo(n)^{-1} \hat w(n)e^{inx}
\nonumber \\
&=& \sum_{n\ne0}
\begin{pmatrix}0 & in\cr \bt^2_1(n)/in &0\end{pmatrix}
\begin{pmatrix}\hat w_1(n)\cr \hat w_2(n)\end{pmatrix}e^{inx}
\nonumber \\
&=& \sum_{n\ne0}
\begin{pmatrix}0 & in\cr \left(\frac{\bt_1(n)}{in}\right)^2 in &0\end{pmatrix}
\begin{pmatrix}\hat w_1(n)\cr \hat w_2(n)\end{pmatrix}e^{inx}.
\label{5-Arep}.
\end{eqnarray}
Thus we have that $w=\Ronetwo U$ satisfies an equation of the form
\eqref{eq:wave} in which $g=\varphi_3$ and
$S$ is a pseudodifferential operator of degree zero defined by
\be
\label{eq:Shat}
\widehat{Sw_1}(n)=
\left(\frac{\bt_1(n)}{in}\right)^2 \hat w_1(n)
\ee
for $n\ne0$. We observe that the eigenvalues of $\Aonetwo$ have the form $\pm\bt_1(k)$. On the other hand, since $\phd$ takes values in $X_1$ and satisfies $\phd(0)=0$, $D\phd(0)=0$, we find that $g(w_1,w_2)=\varphi_3(w)$ is Lipschitz from
a small ball in $\HHone$ into $\IH_1$ with $g(0,0)=0$, $Dg(0,0)=0$, as desired.
\qed

\section{\bf Appendix}

\section{\bf Center manifolds of infinite dimension and codimension}

In this appendix we consider  an abstract differential equations of the form
\begin{equation} \label{eq:FODS}
\frac{du}{dy}(y) = Au(y) + f(u(y)).
\end{equation}
where $X$ and $H$ are Banach spaces with $X$ densely embedded in $H$, $A\in \IL(X,H)$, the space of bounded linear operators from $X$ to $H$, and $f$ is continuously differentiable from $H$ into $H$ with $f(0)=0$ and $Df(0)=0$. We will assume that the nonlinear part $f$ has the regularizing effect: $f(H) \subset X$. This hypothesis on $f$ is a stronger condition than those imposed in many  works related with the existence center manifolds, but this completely natural for the $abcd$-Boussinesq system considered in this paper. Existence of a local infinite dimensional center manifold  was established by J. Quintero and R. Pego  in the case of having a center space with infinite dimension and infinite codimension,  under the hypotheses that the nonlinearity $f$ has a smoothing property described above. We note that the cutoff  performed to get the local manifold must be done  in the $H$ norm, and not in the $X$ norm. This is important since we need to use an energy functional which is defined on $H$, which is conserved in time for classical solutions (taking values in $X$), but is indefinite in general.

We will state the result for the existence of a locally invariant center manifold of classical solutions for the system (\ref{eq:FODS}) under certain conditions which allow the center subspace (that associated with the purely imaginary spectrum of $A$) to have finite dimension and infinite codimension. We start with some basic definitions and some hypotheses:

\begin{Def} Let $J \subset \R$ be an open interval and
$ u\colon {{\R}} \to H $ be a function. We say that $u$ is a classical
solution of \eqref{eq:FODS} on $J$ if the mapping $t \mapsto u(t)$ is
continuous from $ J$ into $ X$, is differentiable from $J$ into $H$
and \eqref{eq:FODS} holds for all $t\in J$.
\end{Def}
Let $\beta>0$, let  $Y$ and $Z$ be Banach spaces and $U$ be an open set in $Y$.
We define the Banach spaces $C_{b}(U,Z)$, $\Lip(U,Z)$ and $Y^\beta$ by
\begin{eqnarray}
C_{b}(U,Z) &:=& \left\{ f \in C(U,Z): \sup_{u\in Z}\|f(u)\|_{Z} < \infty  \right\}. \nonumber \\
\Lip(U,Z) &:=& \left\{ f \in C(U,Z): \|f(u)-f(v)\|_{Z} \leq M_f \|u - v\|_{Y}
\ \mbox{for all} \ u, v \in U \right\}. \nonumber \\
Y^{\beta} &:=& \{u \in C({{\R}},Y): ||u||_{Y^\beta}:=\sup_{t}
e^{- \beta |t|} \|u(t)\|_{Y} < \infty \} \label{Ybeta}.
\end{eqnarray}

Throughout this section we assume that there are bounded
projections $\pi_{0}$ and $\pi_{1}$ on $H$ such that
(i) $H = H_0 \oplus H_1$ with $H_i := \pi_i(H)$,
(ii) $\pi_{i}|_{X}$ is bounded from $X$ to $X$, and
(iii) $AX_i \subseteq H_i$ where $X_i := \pi_i(X) $, for $i=0, 1$.
We let $\pibd$ denote a common norm bound for the projections $\pi_j$
on $H$ and $X$, $j=0, 1$.

As a consequence, the equation \eqref{eq:FODS} can be rewritten as the
first order system
\be\label{eq:Asys}
\begin{array}{rll}
\displaystyle{\frac{d}{dt}}u_0(t) & = & A_0u_0(t) + \pi_0 f(u(t)), \\[8pt] 
\displaystyle{\frac{d}{dt}}u_1(t) & = & A_1u_1(t) + \pi_1 f(u(t)),
\end{array}
\ee
where  $A_i \in \IL(X_i, H_i)$ with $A_i y = \pi_i Ay$ for $y \in X_i$.

We assume the following splitting properties for the operator $A$,
associated with the linear evolution equation $du/dt=Au$.

\begin{itemize}
\item[\bf{(H0)}]
$A_0$ is the generator of a $C^0$-group $\{\Sc(t)\}_{t \in {\R}}$ on
$H_0$ with subexponential growth. I.e.,
given any $\beta >0$, there is a constant $M >0$ such that
\[
 \|\Sc(t)\|_{\IL(H_0)} \leq M  e^{ \beta|t|} \quad \Forall t\in {\R}.
\]

\item[\bf{(H1)}] There exists $\varepsilon>0$ and a positive
function $M_1$ on $[0,\varepsilon)$ such that for any $\varrho\in[0,\varepsilon)$
and for any $g_1\in C(\R,X_1)\cap H_1^\varrho$ the equation
\be\label{eq:hyp}
\frac{d}{dt}u_1 = A_1u_1+g_1
\ee
has a unique solution in $H_1^\varrho$ given by $u_1=K_1g_1$, where
$K_1\in \IL(H_1^\varrho)$ with $\|K_1\|_{\IL(H_1^\varrho)}\le M_1(\varrho)$.
Furthermore $\|K_1\|_{\IL(X_1^\varrho)}\le M_1(\varrho)$.
\end{itemize}
Under those conditions, J. Quintero and R. Pego in (\cite{PQ2}) obtained the following result:
\begin{theorem}[{\bf Local Center Manifold Theorem}] \label{T:1}
Let $H$, $ X$, $A$, $\pi_0$, $\pi_1$ and $f$ be as above,
and let
\[
B(\delta) =\{y\in H_0 : \|y\|_{H}< \delta \}.
\]
Then for all sufficiently small $\delta >0$ there exists
$\phd \colon H_0\to X_1$ such that
\begin{itemize}

\item[(i)] $\phd(0)=0$ and $D\phd(0)=0$.

\item[(ii)]
$\phd\in C_b(H_0,X_1)\cap \Lip(H_0, X_1)$, and on any ball
$B(\delta')$, $\phd$
has Lipschitz constant $L(\delta')$ satisfying $L(\delta')<\frac12$
and $L(\delta')\to0$ as $\delta'\to0^+$.

\item[(iii)] The manifold $\md \subset X$
given by 
\be\label{eq:md}
\md := \{\xi + \phd(\xi): \xi\in X_0 \}
\ee
is a local integral manifold for \eqref{eq:FODS} over
$B(\delta)\cap X_0$.
That is, given any
$y\in \md$ there is a continuous map $u \colon \R \to \md$ with
$u(0)=y$, such that for any open interval $J$ containing $0$ with
$\pi_0 u(J) \subset B(\delta)$ it follows that
$u$ is a classical solution of \eqref{eq:FODS} on $J$.
Moreover, $u_0:=\pi_0u$ is the unique classical solution on $J$
with $u_0(0)=\pi_0y$ to the reduced equation
\be\label{eq:u0}
 \frac{d}{dt}u_0(t) = A_0u_0(t) + \Fd(u_0(t)),
\ee
where $\Fd\colon H_0\to X_0$ is locally Lipschitz and
is given by $\Fd(w):=\pi_0 f(w+\phd(w))$.

\item[(iv)] For any open interval $J\subset\R$,
every classical solution $u_0\in C^1(J,H_0)\cap C(J,X_0)$
of the reduced equation \eqref{eq:u0}
such that $u_0(t) \in B(\delta)$ for all $t \in J$
yields, via $u=u_0 + \phd(u_0)$,  a classical solution $u$
of the full equation \eqref{eq:FODS} on $J$.

\item[(v)] The manifold $\md$ contains all classical solutions on $\R$
that satisfy
$\|u(t)\|_{H} \leq \delta$ for all $t$.
\end{itemize}
\end{theorem}

\nd {\bf Acknowledgments.} J. Quintero was supported by  the Mathematics Department at  Universidad
del Valle C.I. 71007 (Colombia). A. Montes was  supported by  the Mathematics Department at  Universidad del Cauca (Colombia). JRQ and AMP were suported by Colciencias under the research project No 42878.

\end{document}